\documentclass[superscriptaddress]{revtex4}
\usepackage{graphicx,amssymb,epsfig,psfrag}
\usepackage{amsmath}
\usepackage{amsfonts}
\usepackage{amssymb}
\usepackage{color}
\usepackage{subfigure}

\usepackage[utf8x]{inputenc}
\usepackage{ucs}
\usepackage{mathrsfs}

\newcommand{\fig}{Figure }
\newcommand{\change}[1]{{\color{black} #1}}
\newcommand{\xip}{x_i^+}
\newcommand{\xim}{x_i^-}

\newcommand{\tiw}{\tilde{w}}

\newcommand{\tiq}{\tilde{q}}
\newcommand{\tik}{\tilde{k}}
\newcommand{\tio}{\tilde{\omega}}
\newcommand{\tiv}{\tilde{v}}

\newcommand{\dd}{\mathrm{d}}
\newcommand{\ee}{\mathrm{e}}
\newcommand{\ci}{\mathrm{i}}

\let\grad\nabla

\let\grad\nabla

\newcommand{\pard}[2]{\frac{\partial #1}{\partial #2}}

\newcommand{\Real}{\mbox{Re}}
\newcommand{\Imag}{\mbox{Im}}

\begin{document}

\title{\change{Piezoelectric coupling in energy-harvesting fluttering flexible plates : linear stability analysis and conversion efficiency}}

\author{Olivier Doar\'e}
\email{olivier.doare@ensta.fr}
\affiliation{ENSTA, Paristech, Unit\'e de M\'ecanique (UME), Chemin de la Huni\`ere, 91761 Palaiseau, France}
\author{S\'ebastien Michelin}
\email{sebastien.michelin@ladhyx.polytechnique.fr}
\affiliation{LadHyX -- D\'epartement de M\'ecanique, Ecole Polytechnique, Route de Saclay, 91128 Palaiseau, France}
\date{\today}

\begin{abstract}
This paper investigates the energy harvested from the flutter of a plate in an axial flow by making use of piezoelectric materials. The equations for fully-coupled linear dynamics of the fluid-solid and electrical systems are derived. The continuous limit is then considered, when the characteristic length of the plate's deformations is large compared to the piezoelectric patches' length.  The linear stability analysis of the coupled system is addressed from both a local and global point of view. Piezoelectric energy harvesting adds rigidity and damping on the motion of the flexible plate, and destabilization by dissipation is observed for negative energy waves propagating in the medium. This result is confirmed in the global analysis of fluttering modes of a finite-length plate. It is finally observed that waves or modes destabilized by piezoelectric coupling maximize the energy conversion efficiency.
\end{abstract}
\maketitle
%\begin{keyword}
%Fluid-structure interaction, plate, axial flow, flutter instability, energy harvesting, piezoelectricity, negative energy waves
%\end{keyword}
%\end{frontmatter}

%%%%%%%%%%%%%%%%%%%%%%
\section{Introduction}
%%%%%%%%%%%%%%%%%%%%%%

\noindent The environmental impact and limited resources of fossile fuel energies have motivated a significant research effort in the development of new and diverse techniques for the production of electrical energy. In parallel, a particular attention has also been given to systems able to produce limited amounts of energy at low cost to power remote or isolated devices, for which connection to the traditional electrical network is prohibitive in terms of cost or technical complexity \citep{sodano2004}. These two elements have increased the attention on mechanisms able to produce self-sustained vibrations of a solid substrate on one hand and to convert the corresponding mechanical energy into electrical power on the other.

The conversion into electricity of kinetic energy from geophysical flows such as tidal currents, winds and river flows is particularly attractive because of the large availability worldwide and the low environmental impact of this energy source \citep{westwood2004}. Research on fluid-solid interactions has identified several instability mechanisms that can lead to self-sustained vibrations of a solid placed in a steady uniform flow. Such fundamental instability mechanisms as coupled-mode flutter of a heaving and pitching airfoil \citep{mckinney1981}, vortex-induced vibrations \citep{bernitsas2008} or transverse galloping of flexibly mounted structures \citep{barrero2010} are at the core of prototypes or concepts of flow energy harvesters. 

The harvesting of flow energy through flapping of thin elastic plates has also been investigated, mainly using two fundamentally different configurations. In the first one, an unsteady flow, created by the oscillatory wake of an upstream obstacle applies an unsteady forcing on the plate to make it flap \citep{taylor2001,allen2001}. The second configuration uses the coupled-mode flutter instability of the flexible plate in a steady flow  to generate self-sustained flapping \citep{tang2009b}. In that case, it is well-known that the flat equilibrium state of the plate becomes unstable above a critical flow velocity, beyond which dynamic vibrations of large amplitude can develop on the structure. The linear stability of this system has been extensively considered from both a local and a global point of view. In the former, the instability of waves in the infinite medium was considered \citep{brazier1984,crighton1991}, while the latter considered finite systems, including effects such as vortex shedding downstream of the plate \citep{kornecki1976,alben2008d}, three-dimensional effects \citep{eloy2007}, lateral confinement \citep{guo2000}, spanwise confinement \citep{doare2010c} or coupling between multiple structures \citep{michelin2009d}. The non-linear self-sustained flapping developing above the instability threshold has also been the focus of multiple experimental \citep{zhang2000,shelley2005} and numerical studies \citep{alben2008,michelin2008,connell2007}. 

In this work, we are interested in the ability to produce electrical power from the self-sustained oscillations of a flexible plate resulting from this fluttering instability. To assess the potential for electrical energy production, it is important to properly include in the dynamical equations of the fluid-solid sytem the loss of energy due to the conversion into electricity, as we seek here regimes where the extracted energy is a significant fraction of the total energy of the system. Two main approaches can be considered to produce electricity from the mechanical energy of a vibrating solid. Classical generators convert a displacement of a solid substrate into electrical energy through electromagnetic induction, and are commonly used in classical turbines as well as in recent prototypes of flow energy harvesters \citep{bernitsas2008}. On the other hand, piezo-electric materials convert mechanical strain into electric potential, and have recently received increasing attention for applications involving low power production, typically of the order of the mW \citep{sodano2004,anton2007}. Studies on piezoelectric materials for energy harvesting considered flow-induced vibrations \citep{taylor2001,pobering2004,wang2010} but also vibrations from various other forcings such as human movements \citep{platt2005b,shenck2001}.

The objective of the present work is to study from a theoretical point of view the stability properties and dynamics of a classical fluid-solid system (a flexible plate subject to coupled-mode flutter) coupled to an output electrical network with piezoelectric materials. In comparison to previous studies on flow energy harvesting \citep{tang2009b} or piezoelectric energy conversion, the present approach is original by its full coupling of the fluid-solid and electrical systems. In this paper, the linear stability of the fluid-solid-electric system is investigated from both a local and global point of view to identify the impact of the coupling on the stability of the system and assess the efficiency of the mechanical-to-electrical energy conversion. 

Coupling the vibrations with electrical circuits that dissipate energy intuitively results in damping from the point of view of the structure. In the context of fluid-structure interactions, the effect of viscous or viscoelastic damping has been addressed both on the infinite length medium (local approach) and the finite length systems (global approach). \cite{crighton1991} investigated the effect of damping on the stability of flexural waves propagating in compliant panels interacting with an homogeneous flow and found that dissipation can destabilize some particular waves, which are referred to as Negative Energy Waves after \cite{landahl1962}. In the finite length case, the most studied system with respect to the effect of damping is the fluid-conveying pipe, which shares many similarities with the plate in axial flow \citep{paidoussisbook1998,paidoussis2008}. The effect of piezoelectric coupling on the instability threshold of a cantilevered fluid-conveying pipe has been addressed by \cite{elvin2009}. Destabilization or stabilization by dissipation has been observed, depending on the fluid-solid mass-ratio. The comparison of local and global instability criteria with damping has been addressed by \cite{doare2010a}. It was shown that global instability of the long system is always predicted by the local instability criterion of the dissipative medium.

The particular effect of damping induced by piezoelectric coupling has been widely investigated in the research field of structural damping and vibration control. Passive damping by the use of shunted piezoelectric patches, developed by \cite{hagood1991}, has inspired studies involving piezoelectric materials and passive electrical components arranged in a network \citep{maurini2004,bisegna2006}. We will consider here the continuous limit addressed in these two studies, that is valid when the length of the piezoelectric material is small compared to the typical wavelength of the solid's deformations.

In Section \ref{sec:problemformulation}, the linearized equations of motion for the coupled piezo-mechanical problem of a flexible plate covered with piezo-electric elements in a potential flow will be presented. In Sections \ref{sec:local} and \ref{sec:global}, the linear stability analysis will be investigated from a local and global point of view, respectively. In both cases, the effect of piezoelectric coupling on stability will be investigated, as well as the efficiency of energy conversion between the solid mechanical energy and the energy dissipated in a simple resistive electrical network. Finally, in Section \ref{sec:discussion}, the local and global stability results will be discussed as well as possible extensions of the present work.

%%%%%%%%%%%%%%%%%%%%%%%%%%%%%
\section{Problem formulation}
\label{sec:problemformulation}
\begin{figure}
\centering
\begin{psfrags}
\psfrag{PSx}{$x$}
\psfrag{PSxi}{$x_i$}
\psfrag{PSxip}{$x_i^+$}
\psfrag{PSxim}{$x_i^-$}
\psfrag{PSxip1}{$x_{i+1}$}
\psfrag{PSxim1}{$x_{i-1}$}
\psfrag{PSU}{$U_{\infty}$}
\psfrag{PSV1}{$V_i^{(1)}$}
\psfrag{PSV2}{$V_i^{(2)}$}
\psfrag{PSVb}{$\bar{V}_i$}
\psfrag{PSU}{$U_{\infty}$}
\epsfig{file=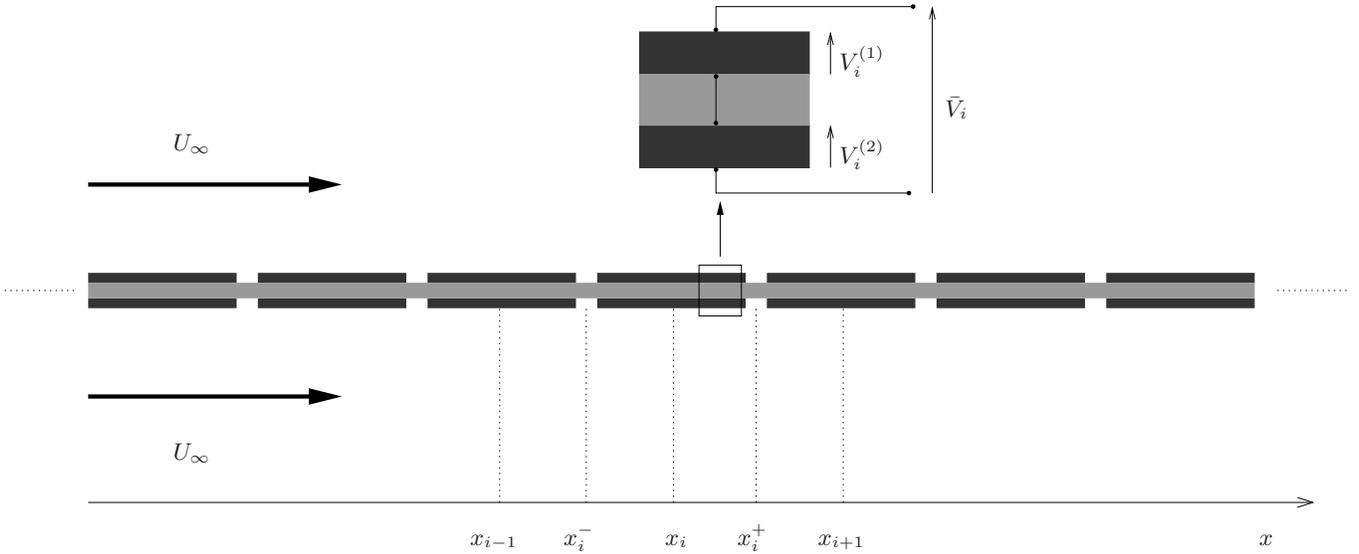,width=\textwidth}
\end{psfrags}
\caption{Schematic view of a plate in an homogeneous axial flow, equipped with small length piezoelectric patches on both sides.\label{fig:sketchdrapeaupiezos}}
\end{figure}

%%%%%%%%%%%%%%%%%%%%%%%%%%%%%
%%%%%%%%%%%%%%%%%%%%%%%%%%%%%
\noindent We consider here the motion of a flexible plate of thickness $h_0$, Young's modulus $E_0$, density $\rho_0$ and Poisson's coefficient $\nu_0$. The span of the flexible plate is assumed to be much greater than its typical streamwise lengthscale. We therefore focus on purely two-dimensional deformations, and the plate's vertical displacement is noted $w(x,t)$. Piezoelectric patches of length $l$, thickness $h_p$, density $\rho_p$, electric permeability $\varepsilon$, Young's modulus $E_p$ and Poisson's coefficient $\nu_p$ are attached symmetrically on the plate. The plate is surrounded by a fluid of density \change{$\rho_f$} with upstream horizontal velocity $U_\infty$. The problem is sketched in Figure \ref{fig:sketchdrapeaupiezos}. In this section, we present the linearized coupled equations of the fluid-solid and piezo-electric systems, first in the case of discrete piezoelectric patches, and then in the continuous limit when $l$ is much smaller than the streamwise lengthscale of the solid deformation. In this limit, the equation of energy conservation will enable us to exhibit the different energy transfers in the system. In the following, for a function $a(x,t)$, $\dot{a}$ and $a'$ correspond to the temporal and streamwise derivatives of $a$ respectively.

% ------------------------------
% ------------------------------

%%%%%%%%%%%%%%%%%%%%%%%%%%%%%%%%%%%%%%%%%%%%%%%%%%%%%%%%%%%%%%%%%%%%%%%%%%%%%%%%%%%%%%%%%%%%%%%%%%%%
\subsection{\change{Equilibrium equations of a beam with discrete pairs of piezoelectric patches}}
%%%%%%%%%%%%%%%%%%%%%%%%%%%%%%%%%%%%%%%%%%%%%%%%%%%%%%%%%%%%%%%%%%%%%%%%%%%%%%%%%%%%%%%%%%%%%%%%%%%%

The left and right ends of the $i^{th}$ piezoelectric pair are positioned in $\xim$ and $\xip$, so that $l=\xip-\xim$ and $x_i=(\xim+\xip)/2$ denotes the position of the center of the patch. We consider complete coverage of the plate by piezoelectric patches, therefore $x_i^+=x_{i+1}^-$. The following equations can easily be adapted to the case of partial coverage \citep{bisegna2006}. Quantities related to the piezoelectric patch located on the upper (respectively, lower) face are denoted by the exponent $^{(1)}$ (respectively, $^{(2)}$). The patches are attached to the plate so that their respective polarities are reversed, and the charge displacement across a piezoelectric patch is obtained as \citep{preumont2002,thomas2009}: 
\begin{equation}
Q_i^{(k)} = \chi \Big[ w' \Big]_{\xim}^{\xip}+C V_i^{(k)}, \;\;\; k=1,2,
\label{eq:chargediscrete}
\end{equation}
where $\chi= e_{31} (h_0+h_p)/2$ is a mechanical/piezoelectrical conversion factor with $e_{31}$ the coupling factor, and $C=\varepsilon\,l/h_p$ is the capacity per unit length in the spanwise direction of the piezoelectric element. $V_i^{(k)}$ is the potential difference between the two electrodes of the corresponding piezoelectric (see \fig \ref{fig:sketchdrapeaupiezos}). Negative electrodes of the patches are assumed to be shunted through the plate by a purely conductive material, therefore,
\begin{equation}
Q_i^{(1)}=Q_i^{(2)}=Q_i,
\end{equation}
and
\begin{equation}
V_i^{(1)}=V_i^{(2)}=V_i.
\end{equation}
Each piezoelectric pair is therefore equivalent to a single piezoelectric patch of equivalent capacity $\bar{C}$ and voltage $\bar{V}_i$ of respective values,
\begin{equation}
\bar{C} = \frac{C}{2}, \;\;\; \bar{V}_i = 2 V_i,
\end{equation}
and this representation is retained for simplicity in the remaining of the paper. Assuming an Euler--Bernoulli model for the plate, the bending moment at a given position $x$  results from the internal rigidity of the material and the piezoelectric coupling:
\begin{equation}
\mathcal{M} = B w'' - \sum_i \chi \bar{V}_i F_i,
\label{eq:momentum}
\end{equation}
where $B$ is the flexural rigidity of the three-layer plate \citep{lee1989},
\begin{equation}\label{eq:rigidity}
B = \frac{E_0 h_0^3}{12(1-\nu_0^2)} + \frac{2 E_p h_p }{1-\nu_p^2} \left( \frac{h_0^2}{4} + \frac{h_0h_p}{2} + \frac{h_p^2}{3} \right),
\end{equation}
and $F_i$ is the characteristic function of the $i^{th}$ patch 
\begin{equation}
F_i = H(x-\xim)-H(x-\xip).
\end{equation}
with $H$ the Heavyside step function. \change{In Eq.~\eqref{eq:momentum}, the summation over all piezoelectric patches and the use of the characteristic functions $F_i$ provides a compact expression for the local bending moment induced by the piezoelectric coupling. }The linearized conservation of momentum for the plate then leads to
\begin{equation}
\mu \ddot{w} = -\mathcal{M}'' - [P],
\label{eq:localeq1}
\end{equation}
where $\mu$ is the surface density of the plate with piezoelectric elements and $[P]=P^+-P^-$ is the pressure forcing applied by the fluid on the plate. We consider here a purely inviscid model for the flow, and the exact expression of $[P]$ will be made explicit in the following sections for the infinite- and finite-length systems respectively.  \change{In non-inviscid flow, the viscous boundary layers would tend to stabilize the plate by inducing a non-uniform tension, maximum upstream \citep[see for example][]{connell2007}. For large enough Reynolds numbers, this correction is however expected to be small. }

\begin{figure}
\centering
\begin{psfrags}
\psfrag{PSG}{$1/G$}
\psfrag{PSE}{$E$}
\psfrag{PSC}{$\tilde{C}$}
\psfrag{PSV}{$\tilde{V}$}
\epsfig{file=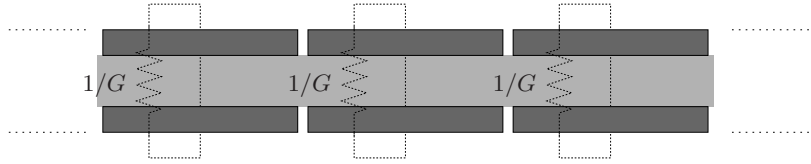,width=0.6\textwidth}
\caption{All piezoelectric pairs are shunted with resistances, modeling the electrical energy absorption. \label{fig:shunts}}
\end{psfrags}
\end{figure}

An additional relation between $\bar{V}_i$ and $Q_i$ is needed to close the system of equations \eqref{eq:chargediscrete}, \eqref{eq:momentum} and \eqref{eq:localeq1}, and is obtained from the output electrical network connected to the free electrodes of the piezoelectric patches.
\cite{lesieutre2004} showed that energy removal from the piezoelectric system induces structural dissipation, whatever electric device is effectively used (resistor, storage in a battery or other energy harvesting circuitry). Consequently, the simplest electrical component, where each piezoelectric pair is shunted by a conductance $G$, is considered to investigate the potential harvesting power of the system and its effect on the flutter dynamics (\fig \ref{fig:shunts}). Applying Ohm's law,
\begin{equation}\label{eq:electricdiscrete}
\bar{V}_i + \frac{\dot{Q}_i}{G} = 0.
\end{equation}
This simple electric network is a particular case of more general electric networks considered in \cite{bisegna2006}, in the context of the optimization of dynamical properties of pinned-pinned beams. 

%%%%%%%%%%%%%%%%%%%%%%%%%%%%%
\subsection{Continuous limit}
%%%%%%%%%%%%%%%%%%%%%%%%%%%%%
\noindent If the typical lengthscale of the plate's streamwise deformations is large compared to the length $l$ of the piezoelectric patches, one may consider the continuous limit of the discrete equations \eqref{eq:chargediscrete}, \eqref{eq:momentum}, \eqref{eq:localeq1} and \eqref{eq:electricdiscrete}. We introduce the surface density of the charge, piezoelectric capacity and conductance of the resistive circuit as
\begin{equation}
q(x_i) = Q_i/l, \;\; c = \bar{C}/l, \;\; g = G / l,
\end{equation}
and the continuous voltage $v(x=x_i)=\bar{V}_i$. As $l\rightarrow 0$,
\begin{align}
\Big[ w' \Big]_{\xim}^{\xip}
	& \simeq w''(x_i) l,  \\ %+ \mathcal{O}(dx^3), \\
\sum_i \bar{V}_i F_i(x)& \simeq v(x).
%\delta'(x-\xim) - \delta'(x-\xip) & \simeq \delta''(x-x_i) \; l , \\ % + \mathcal{O}(dx^3). %, \\
%2V_i - V_{i-1} - V_{i+1} & = \frac{\partial^2 V_i}{\partial x^2} dx^2 + \mathcal{O}(dx^3).
\end{align}
%and the sum appearing in Eq.~\eqref{eq:poutrepiezodiscrete} then takes the form of an integral when $l\rightarrow 0$,
%\emph{(Ici pour etre coherent il faudrait peut etre mettre l'expression avec les O(dx)? De plus, il me semble qu'il y a une incoherence: a gauche on a une sommation sur l'indice i, alors qu'a droite on n'a qu'une seule valeur?)}
%\begin{align}
%\sum_i \bar{V}_i \left[ \delta'(x-\xim) - \delta'(x-\xip) \right]
%& \simeq \sum_{i \in \texttt{Z}} v(x_i) \delta''(x-x_i) l \\
%& \simeq \int_{-\infty}^{+\infty} v(\xi) \delta''(x-\xi) d\xi \\
%& \simeq v''(x).
%\end{align}
%Keeping only the dominant terms in Eqs.~\eqref{eq:chargediscrete}, \eqref{eq:poutrepiezodiscrete} and \eqref{eq:electricdiscrete} when $dx\rightarrow 0$, the continuous equations are obtained as
The continuous equations are then obtained as
\begin{align}
\mu \ddot{w} + B w'''' - \chi v'' & = -[P] \label{eq:coupledw1},\\
q - c v - \chi w'' & = 0 \label{eq:coupledq1}, \\
F(q,v)  &= 0.
\end{align}
Here, $F(q,v)$ is a general expression that links charge and voltage, so that at this stage, the equations are valid for any circuit. In the particular case considered here, $F(q,v)=v+\dot{q}/g$ and $v$ can be eliminated from the previous dynamical equations, thereby leading to a system for $w$ and $q$ only:
\begin{align}
\left( B+\frac{\chi^2}{c} \right) w'''' + \mu \ddot{w} - \frac{\chi}{c} q'' &= -[P] \label{eq:coupledw2},\\
\frac{c}{g}\dot{q} + q - \chi w'' &= 0 \label{eq:coupledq2}.
\end{align}
These continuous equations are similar to those obtained in the Laplace space by \cite{bisegna2006} where homogenization techniques were used. %Here, it is written in the form of local wave equations, suitable for wave propagation and stability analysis.

%%%%%%%%%%%%%%%%%%%%%%%%%%%%%%%%%%%%%%%%%%%%%%%%%%%%%%%%%%%%%%%%%%
\subsection{Efficiency of the energy conversion}
%%%%%%%%%%%%%%%%%%%%%%%%%%%%%%%%%%%%%%%%%%%%%%%%%%%%%%%%%%%%%%%%%%
\noindent To assess the amount of electrical power that can effectively be extracted, nonlinear effects are important to provide the saturation amplitude of the self-sustained oscillations. This question is not addressed here as we focus only on the linear analysis, and will be the subject of a subsequent contribution. However, important physical insight can be gained from the analysis of energy transfer by linearly unstable modes. \change{In particular, the non-linear mode shape that determines the piezoelectric deformation rate and the energy transfers to the output circuit, has been observed to be similar to that of the most linearly unstable mode \citep{eloy2008,michelin2008}.} From Eqs.~\eqref{eq:coupledw1} and \eqref{eq:coupledq1}, the equation for energy conservation can be obtained as

\begin{equation}\label{eq:energy2}
\pard{\mathscr{E}}{t}=\mathscr{P}_p-\mathscr{P}_{el}+\pard{\mathscr{F}}{x},
\end{equation}
with 
\begin{align}
\mathscr{E}&=\frac{1}{2}\rho_s \dot{w}^2+\frac{1}{2}Bw^{''2}+\frac{1}{2}cv^2,\\
\mathscr{F}&=\dot{w}\left(\chi v'-Bw'''\right)+\dot{w}'\left(Bw''-\chi v\right),\\
\mathscr{P}_p&=-p\dot{w},\\
\mathscr{P}_{el}&=-v\dot{q}.
\end{align}
The total energy density of the system $\mathscr{E}$ is the sum of the solid kinetic and elastic energy as well as the electrical energy stored in the capacity of the piezoelectric material. \change{$\mathscr{F}$ is the flux of mechanical energy in the plate: the first and second terms are respectively the rate of work of the internal bending forces and moments, due both to the elastic response of the solid and the piezoelectric coupling.} $\mathscr{P}_p$ is the rate of work of the local pressure force and $\mathscr{P}_{el}$ is the power readily available and dissipated in the output system. Note that this equation is valid regardless of the output electrical network chosen. In the case of the purely resistive system considered here, $\mathscr{P}_{el}=\dot{q}^2/g$.

%%%
% J'ai viré le truc suivant parce que je trouve qu'on se répète un peu quand-meme.
%%%
%To define the complete efficiency of the energy transfer between the fluid and the resistive circuit, a non-linear approach is necessary. Instead, here,
We consider a measure of the harvested (i.e. dissipated) energy over one flapping period $T$ relative to the mean energy density of the solid-piezoelectric system during that period. In that regard, this ratio is a normalization of the harvested energy, and it will be referred to as conversion efficiency in the following. This ratio is defined for an unstable mode (that can lead to self-sustained oscillations) as
\begin{equation}
r = \frac{\displaystyle\int_0^{T} \langle \mathscr{P}_{el} \rangle dt }{\displaystyle\frac{1}{T}\int_0^{T} \langle \mathscr{E} \rangle dt}.
\label{eq:efficiency}
\end{equation}
In this last expression $\langle . \rangle$ stands for the spatial mean value for the considered mode, taken over either a wavelength in the local analysis or the entire plate in the global analysis. Note that since $r$ is just a normalized energy output and not a thermodynamic efficiency, $r>1$ is allowed. In the following, we will study the influence of the system's parameters on this ratio to find optimal conditions for the energy conversion.

%%%%%%%%%%%%%%%%%%%%%%%%%%%%%%%%%%%%%%%%%%%%%%
\section{Destabilization by damping and energy conversion in the infinite medium}
\label{sec:local}
%%%%%%%%%%%%%%%%%%%%%%%%%%%%%%%%%%%%%%%%%%%%%%

%%%%%%%%%%%%%%%%%%%%%%%%%%%%%%%%%%%%%
\subsection{Non-dimensional equations}
%%%%%%%%%%%%%%%%%%%%%%%%%%%%%%%%%%%%%
\noindent Using $\mu/\rho_f$, $\mu/\rho_f U_{\infty}$, $\rho U_{\infty}^2$ and $U_{\infty} \sqrt{\mu c}$ respectively as characteristic length, time, pressure and charge surface density, and noting all non-dimensional variables with a tilde, Eqs.~\eqref{eq:coupledw2} and \eqref{eq:coupledq2} take the following non-dimensional form:
\begin{align}
\frac{1}{V^{*2}}(1+\alpha^2)\tiw'''' + \ddot{\tiw} - \frac{\alpha}{V^*} \tiq'' & = -[\tilde{p}], \label{eq:localadimfluid1}\\
\gamma \dot{\tiq} + \tiq - \frac{\alpha}{V^*} \tiw'' & = 0 \label{eq:localadimfluid2},
\end{align}
with
\begin{align}
V^* &= \sqrt{\frac{\mu^3 U_{\infty}^2}{B \rho_f^2}} \label{eq:Vstar}\\
\alpha &= \frac{\chi}{\sqrt{cB}} \label{eq:alpha}\\
\gamma &= \frac{\rho_f U_{\infty} c}{\mu g}.\label{eq:gamma}
\end{align} 
$V^*$ is the non-dimensional flow velocity, $\alpha$ is the piezoelectric coupling coefficient, and $\gamma$ is the ratio between the fluid-solid and electrical characteristic timescales. %In the remaining of this section, tildes will be omitted on space and time variables, for the sake of clarity.

In the present local analysis, mechanical and electrical displacements are sought in the form of harmonic waves of wavenumber $\tik$ and frequency $\tio$,
\begin{equation}
\left(
\begin{array}{c}
\tiw \\
\tiq
\end{array}
\right)
=
\Real\left[
\left(
\begin{array}{c}
w_0 \\
q_0
\end{array}
\right)
\ee^{\ci(\tilde{k}x- \tilde{\omega} t)}
\right].
\label{eq:depltonde}
\end{equation}

\subsection{Computation of the pressure forcing}
\noindent The pressure forces are computed assuming a potential flow on both sides of the flexible solid. The potential $\Phi$ can be decomposed in $\Phi(x,y,t)=x+\phi(x,y,t)$ where $\phi$ is the perturbation potential to the uniform base flow. The flow is incompressible therefore $\phi$ must satisfy
\begin{equation}
\nabla^2 \phi = 0,
\label{eq:laplace}
\end{equation}
with boundary conditions,
\begin{align}
\pard{\phi}{y} (x,y=0^+,t)=\pard{\phi}{y}(x,y = 0^-)&= \dot{w}(x,t) + w'(x,t),\label{eq:transpbc}\\
\grad\phi\rightarrow 0,\qquad & \textrm{for    } |y|\rightarrow\infty.\label{eq:bcinf}
\end{align}
Note that $\phi$ is discontinuous on the plate but the normal velocity must be continuous. The pressure is obtained from the linearized unsteady Bernoulli equation as
\begin{equation}
\tilde{p} = - \pard{\phi}{t} - \pard{\phi}{x}.
\label{eq:bernoulli}
\end{equation}
The velocity potential is obtained solving Eq.~\eqref{eq:laplace} with boundary conditions \eqref{eq:transpbc}--\eqref{eq:bcinf}. The pressure jump is then obtained by applying Eq.~\eqref{eq:bernoulli} on both sides of the plate. From Eq.~\eqref{eq:depltonde}, it can be written
\begin{equation}
[\tilde{p}] = \tilde{p}(x,y=0^+,t) - \tilde{p}(x,y=0^-,t) = \Real \left[ 2 w_0 \frac{(\tio-\tik)^2}{|\tik|} e^{i(\tik x - \tio t)} \right]\cdot
\label{eq:pressureonde}
\end{equation}
Using Eqs.~\eqref{eq:pressureonde} and \eqref{eq:depltonde}, the system \eqref{eq:localadimfluid1}--\eqref{eq:localadimfluid2} becomes a linear system for $w_0$ and $q_0$,
\begin{equation}
\left[
\begin{array}{cc}
D_0 + D_1^2 & D_1 \\
D_1 & D_2
\end{array}
\right]
\left\{
\begin{array}{c}
w_0 \\
q_0
\end{array}
\right\}
=
\mathcal{L}(\tik,\tio) 
\left\{
\begin{array}{c}
w_0 \\
q_0
\end{array}
\right\}
=
\left\{
\begin{array}{c}
0 \\
0
\end{array}
\right\},
\label{eq:systonde}
\end{equation}
with
\begin{align}
D_0 & = -\tio^2\left(1+\frac{2}{|\tik|}\right) - 2 |\tik| + 4 \tio \frac{\tik}{|\tik|} + \frac{\tik^4}{V^{*2}} \label{eq:D0}, \\
D_1 & =  \frac{\alpha \tik^2}{V^*} \label{eq:D1}, \\
D_2 & = 1 - i \gamma \tio \label{eq:D2}.
\end{align}
Equation \eqref{eq:systonde} admits non trivial solutions if and only if the determinant of $\mathcal{L}$ vanishes, which leads to the following dispersion relation:
\begin{equation}
D_0(\tik,\tio) = D_1(\tik,\tio)^2 \left( \frac{1}{D_2(\tik,\tio)} - 1 \right).
\label{eq:reldispcoupled}
\end{equation}
For $(\tik,\tio)$ solution of Eq.~\eqref{eq:reldispcoupled}, Eq.~\eqref{eq:systonde} determines the relative amplitude of the mechanical displacement and electrical charge for the corresponding wave. 
%%%%%%%%%%%%%%%%%%%%%%%%%%%%%%%%%%%%%
\subsection{Local stability analysis}
%%%%%%%%%%%%%%%%%%%%%%%%%%%%%%%%%%%%%

%\emph{Je pense que cette section peut etre un peu raccourcie, en particulier le premier paragraphe. De plus, je ne suis pas un specialiste de l'analyse locale et des travaux de Crighton mais il faut veiller a ne pas redetailler des calculs deja publies dans le cas non couple, et ne garder que ce qui est strictement necessaire pour le cas couple, qui est la valeur ajoutee de notre papier.}

\noindent We will restrict the wave analysis to the temporal approach, which classically consists in looking for frequencies satisfying the dispersion relation associated to a real wavenumber $\tik$. Equation~\eqref{eq:reldispcoupled} can be put in the form of a third-order polynomial in $\tio$. Hence, for any real wavenumber, there are three different waves. If there exists a real wavenumber for which one of these frequencies has a positive imaginary part, the corresponding wave grows exponentially in time, indicating a temporal instability. In the energy harvesting context, this temporal instability is necessary to create self-sustained oscillations of the plate that are able to generate a net power to the electric networks via the piezoelectric coupling.\\

Let us first consider the case of no-piezoelectric coupling ($\alpha=0$): the matrix $\mathcal{L}$ is diagonal and the problem then consists in two distinct dispersion relations. The first describes the propagation of flexural waves in the medium,
\begin{equation}
D_0(\tik,\tio) = 0.
\label{eq:reldisp_nocoupling1}
\end{equation}
This dispersion relation is very similar to that of a compliant panel interacting with a potential flow, which has been extensively studied \citep{brazier1984,crighton1991}. The difference between the present dispersion relation and that of \cite{brazier1984} comes from the different choice for the non-dimensional time and the presence of the flow on both sides of the plate. Hence, the same phenomena will be observed, but at different values of the non-dimensional velocity, frequencies and wavenumbers. The main feature of this medium is that it is unstable for any non-zero value of the flow velocity. Analyses of the different branches in the complex $\tik$- and $\tio$-planes have also been performed in the above-mentioned papers to investigate the convective-absolute instability transition, which will not be addressed here. %\emph{(Par exemple ici, Shelley et al. (2005) a fait le calcul a la Crighton pour le drapeau donc je pense qu'on peut citer Shelley et zapper une bonne partie de la discussion de $D_0$)}

The second uncoupled dispersion relation describes the behavior of the charge in the electric network,
\begin{equation}
D_2(\tik,\tio)=0.
\label{eq:reldisp_nocoupling2}
\end{equation}
Equation \eqref{eq:reldisp_nocoupling2} corresponds to the charge dynamics in an RC circuit and does not include the wavenumber $\tik$, as no propagation of charge can exist in a medium composed of electrical circuits disconnected from each other. 
% -------------------
\begin{figure}[h]
\centering
\begin{psfrags}
\psfrag{PSkb}{$k_b$}
\psfrag{PSks}{}
\psfrag{PSkp}{}
\psfrag{PSkc}{$k_c$}
\epsfig{file=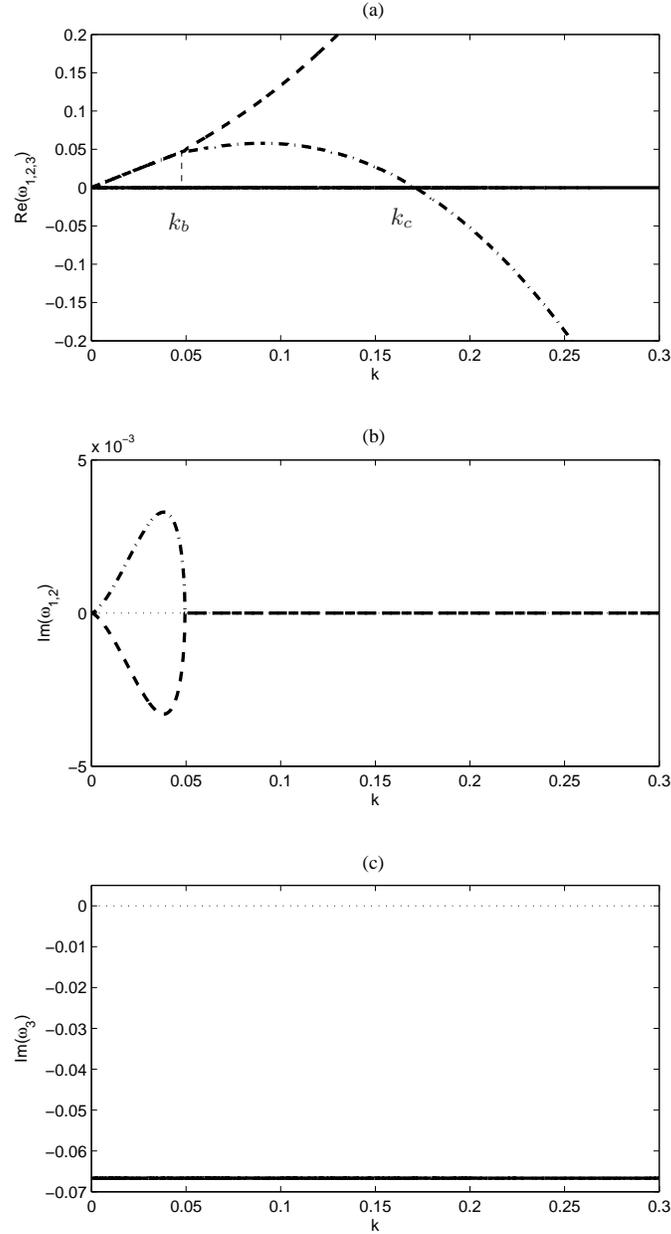,height=0.8\textheight} \\
\end{psfrags}
\caption{Real part (a) and imaginary part (b,c) of frequencies associated to a real wavenumber $k$, for $V^* = 0.05$, $\gamma=15$ and $\alpha=0$. This corresponds to a situation where no coupling is present in the system.\label{fig:branches_nocoupling}}
\end{figure}
% -------------------

On Figure \ref{fig:branches_nocoupling}, the frequencies $\tio_n(\tik)$ ($n=1..3$) are plotted. Here, $\tio_1$ and $\tio_2$ are solutions of Eq.~\eqref{eq:reldisp_nocoupling1} and correspond to flexural waves, while $\tio_3$ is the solution of Eq.~\eqref{eq:reldisp_nocoupling2} and corresponds to an electrical wave. For $\tik \in [0,\tik_b]$, the two frequencies $\tio_1$ and $\tio_2$ are complex conjugate, one of them having a positive imaginary part that indicates a temporal instability. For $\tik>\tik_b$, $\tio_1$ and $\tio_2$ are real and the waves are neutrally stable.  Wave 1 has positive phase velocity for all $k$ but two ranges of wavenumbers can be distinguished depending on the sign of the phase velocity of wave 2. For $\tik_b\leq k\leq\tik_c$, $\tio_2>0$ and wave 2 has a positive phase velocity, while for $k\geq \tik_c$, it has a negative phase velocity. As a consequence of the phase velocity sign change, the frequency $\tio$ vanishes for $\tik=\tik_c$. From Eq.~\eqref{eq:reldisp_nocoupling1}, one can show that
\begin{equation}
\tik_b \sim V^*,
\end{equation}
when $V^{*} \ll 1$. Looking for zeroes of $\tio_2$, we obtain
\begin{equation}
\tik_c=2^{1/3}V^{*2/3}.
\end{equation}
Finally, the frequency $\tio_3$ is constant and purely imaginary with negative imaginary part: no energy propagates and the negative growth rate $\text{Im}(\tio_3)$ is the characteristic time of discharge of a capacity $c$ in a resistance $1/g$.\\

\begin{figure}[h]
\centering
\begin{psfrags}
\psfrag{PSkb}{$k_b$}
\psfrag{PSk}{$k$}
\psfrag{PSE}{$E$}
\psfrag{PSkc}{$k_c$}
\epsfig{file=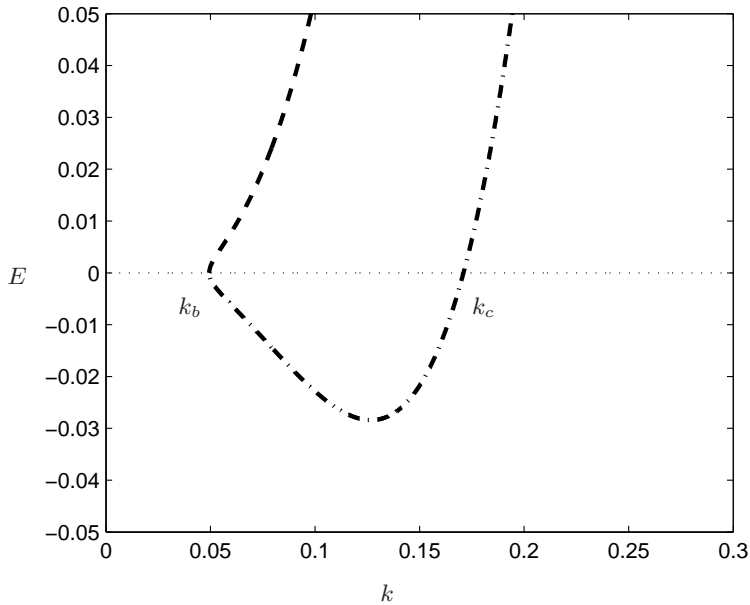,width=0.6\textwidth} \\
\end{psfrags}
\caption{Wave energy of the neutral waves propagating in the system without piezoelectric coupling for  $V^* = 0.05$.\label{fig:waveenergy}}
\end{figure}

\begin{figure}[h]
\centering
\epsfig{file=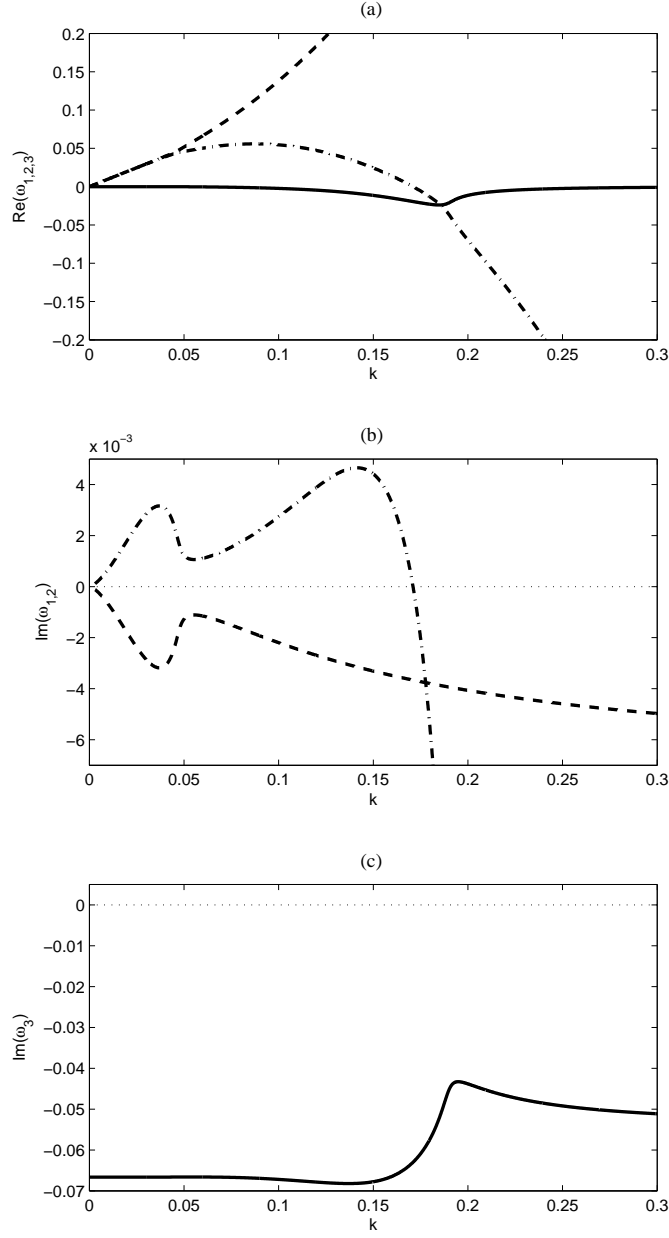,height=0.8\textheight} \\
\caption{Real part (a) and imaginary part (b,c) of frequencies associated to a real wavenumber $k$, for $V^* = 0.05$, $\gamma=15$ and $\alpha=0.5$. This corresponds to a situation where coupling is present in the system.\label{fig:branches_coupling}}
\end{figure}

Piezoelectric coupling is now considered ($\alpha\neq 0$). From a mechanical point of view, its effect is to pump energy from the system to produce heat in the resistances, thereby dissipating mechanical energy. It is thus expected that results regarding the effect of damping on stability of neutral waves will also apply here. In particular, as demonstrated by \cite{landahl1962}, the sign of wave energy is expected to allow the prediction of the stabilizing or destabilizing effect of coupling on neutral waves. The wave's energy is defined as the work done in slowly building up the wave starting from rest at time $t=-\infty$ and has for expression \citep{cairns1979},
\begin{equation}
E = - \frac{\tio}{4} \frac{\partial D_0}{\partial \tio} |w_0|^2.
\label{eq:cairnsenergy}
\end{equation}
Let us consider a neutral wave propagating in the uncoupled system. The corresponding frequency is $\tio_0(\tik)$, so that $D_0(\tio_0,\tik)=0$ and $\tio_{0i}=0$. The perturbed value of $\tio$ due to the addition of piezoelectric coupling is then written as,
\begin{equation}
\tio = \tio_0 + \delta \tio,
\end{equation}
where $\delta \tio \ll \tio$. The frequency $\tio$ satisfies the dispersion relation, thus,
\begin{equation}
\delta \tio \left. \frac{\partial D_0}{\partial \tio}\right|_{(\tik,\tio_0)} \simeq \left( \frac{D_1^2}{D_2} - D_1^2 \right)_{(\tik,\tio_0)}.
\label{eq:reldisp}
\end{equation}
In this last expression, only the leading order terms have been kept. The stabilizing or destabilizing effect of piezoelectric coupling depends on the imaginary part of $\delta \tio$. After a straightforward calculation, this quantity reads,
\begin{equation}
\delta \tio_i \simeq \frac{\tio_0 \alpha^2 \gamma \tik^4}{V^{*2}(1+\tio_0^2\gamma^2)\dfrac{\partial D_0}{\partial \tio}}\cdot
\end{equation}
It hence appears that the variation of the growth rate after the addition of piezoelectric coupling has the opposite sign of the neutral waves energy in absence of coupling. It is positive for a negative energy wave (NEW), and negative for a positive energy wave (PEW). Following the classification of \cite{benjamin1963}, NEW are also referred to as {\em class A} waves, while PEW are referred to as {\em class B} waves. On Figure \ref{fig:waveenergy}, energy of waves 1 and 2 are plotted as function of the wavenumber. Energy of wave 2 is negative for $\tik\in[\tik_b,\tik_c]$ and the wave energy analysis shows that this range of wavenumbers will become unstable when piezoelectric coupling is added. To address the validity of the previous prediction, Figure \ref{fig:branches_coupling} represents the three frequencies associated \change{with} a real wavenumber in the case $V^* = 0.05$, $\alpha=0.5$, so that it is the same case as in Figures \ref{fig:branches_nocoupling} and \ref{fig:waveenergy}, but with added piezoelectric damping. Here, piezoelectricity couples mechanical displacement and electrical displacement. Consequently, no purely mechanical or electrical waves propagate in the system. The predictions of the above wave energy analysis are confirmed by the behavior of $\Imag (\tio_1)$ and $\Imag (\tio_2)$ on Figure \ref{fig:branches_coupling}b: Wave 1 is stabilized by the addition of piezoelectric coupling while wave 2 is destabilized by the addition of piezoelectric coupling in the range of wavenumbers $[\tik_b,\tik_c]$ and stabilized for $\tik>\tik_c$. For the third wave, which is a purely electrical wave without coupling, the frequency has now a non-zero real part, indicating that it is now a propagating wave. Consequently, this wave also appears to be affected by coupling.
%To conclude, the stability properties of waves are affected by the presence of piezoelectric coupling in the same manner as with more classical types of damping. Negative energy waves are destabilized so that the range of unstable waves is $[0, \tik_c]$ with piezoelectric coupling while it was $[0,\tik_b]$ in the uncoupled case.

%%%%%%%%%%%%%%%%%%%%%%%%%%%%%%%%%%%%%%%%%%%%%%%%%%%%%%%%%%%
\subsection{Energy conversion efficiency of unstable waves}
\begin{figure}[h]
\centering
\begin{psfrags}
\psfrag{PSkb}{$k_b$}
\psfrag{PSkm}{$k_m$}
\psfrag{PSkp}{$k_p$}
\psfrag{PSkc}{$k_c$}
\psfrag{PSgamma}{$\gamma$}
\psfrag{PSO}{$\!\!\!\!\!\!\Omega$}
\psfrag{PSK}{$\!\!\!\!\!\!K$}
\psfrag{PSR}{$\!\!\!\!\!\!R$}
\epsfig{file=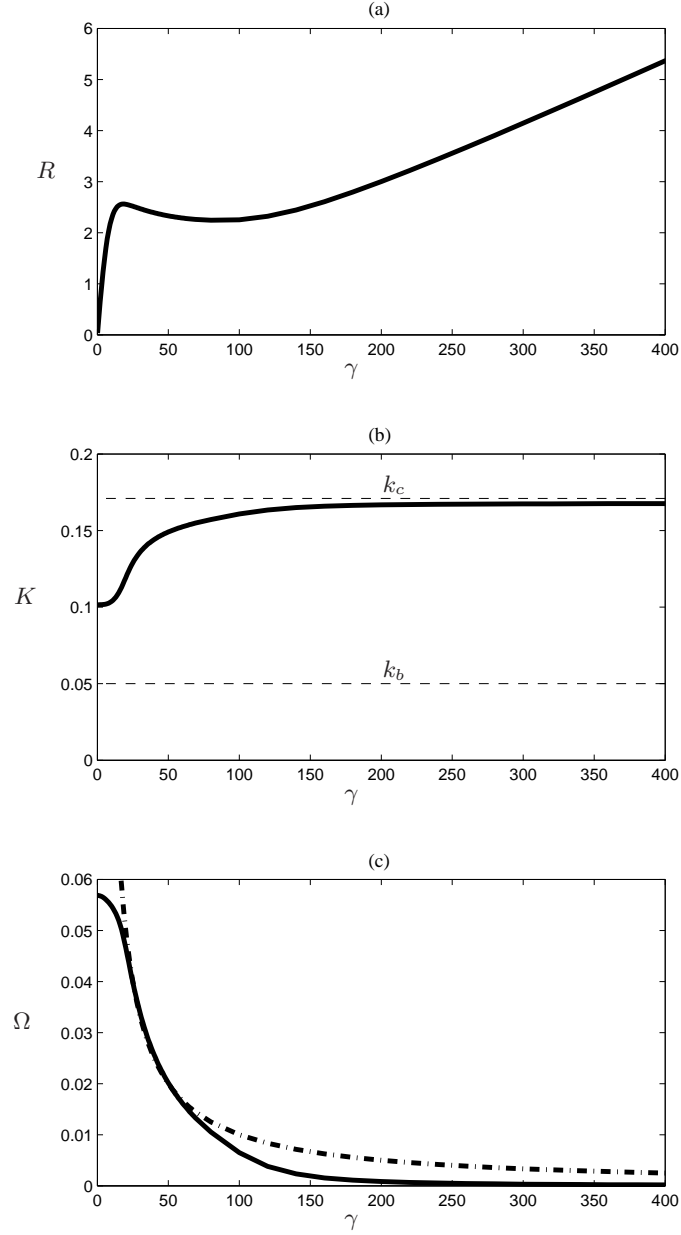,height=0.8\textheight}
\end{psfrags}
\caption{(a) Maximal efficiency, $R$ as function of $\gamma$ for $\alpha=0.5$ and $V^*=0.05$; (b) Value of the wavenumber of the corresponding unstable wave; (c) value of the corresponding frequency, compared with $1/\gamma$ (dashed line). \label{fig:R_gamma}}
\end{figure}

%%%%%%%%%%%%%%%%%%%%%%%%%%%%%%%%%%%%%%%%%%%%%%%%%%%%%%%%%%%
\noindent Now that the effect of piezoelectric coupling on the propagation of waves and their stability has been addressed, the energy conversion efficiency of these waves is investigated. As defined in the previous section, energy conversion is modeled by resistances that shunts the piezoelectric patches, and is significant only if a wave is unstable, so that it grows exponentially in time and eventually saturates through a nonlinear mechanism that is not addressed in the present analysis. The starting point is the ratio defined in Eq.~\eqref{eq:efficiency}, which takes the following form when non-dimensional variables are used:
\begin{equation}
r = \frac{\dfrac{1}{\gamma}\displaystyle\int_0^{2\pi/\tio_r} \langle \tiv \dot{\tiq} \rangle \,\dd t}
{\dfrac{\tio_r}{2\pi}\displaystyle\int_0^{2\pi/\tio_r} \dfrac{1}{2} \langle \dot{\tiw}^2 + \dfrac{1}{V^*} \tiw^{''2} + \tiv^2 \rangle \,\dd t}.
\label{eq:efficiencylocal}
\end{equation}
The spatial averaging appearing in this last expression corresponds to the mean value over one wavelength. When it is applied to the product of two quantities transported by a wave of wavenumber $\tik$ and frequency $\tio$, namely $a=\Real[a_0e^{i(\tik x - \tio t)}]$ and $b=\Real[b_0e^{i(\tik x - \tio t)}]$, it reads
\begin{equation}
\langle a b \rangle = \frac{\tik}{2\pi} \int_0^{2\pi/\tik} ab \,\dd x = 2 e^{2 \tio_i t} \Real(a_0\bar{b}_0).
\end{equation}
The numerator and denominator of Eq.~\eqref{eq:efficiencylocal} have the same time-dependence therefore $r$ does not depend on time. With a displacement in the form of Eq.~\eqref{eq:depltonde}, it finally takes the following form,
\begin{equation}
r(\alpha,V^*,\gamma,\tik,\tio) = \dfrac{8\pi}{\tio_r} \left[ \gamma^{-1} \left( 1 + \dfrac{\tik^4}{V^*|\tio|^2}\right) \left| V^* \frac{i \gamma \tio - 1}{\alpha \tik^2} \right|^2 + \gamma \right]^{-1},
\end{equation}
where $\tio$ and $\tik$ are linked through the relation dispersion Eq.~\eqref{eq:reldispcoupled}. For a given set of parameters $\alpha$, $V^*$ and $\gamma$, let $R$ be the maximum value of $r$ among all unstable waves,
\begin{equation}
R(\alpha,V^*,\gamma) = \max_{\tik, \tio_{i}>0} r(\alpha,V^*,\gamma,\tik,\tio).
\end{equation}
The wavenumber and frequency of the corresponding wave are noted $K$ and $\Omega$, respectively. The maximum efficiency $R$ is plotted in Fig. \ref{fig:R_gamma} as function of $\gamma$, for $\alpha=0.5$ and $V^*=0.05$. It presents a local maximum at $\gamma\sim 22$ and tends to infinity for high values of $\gamma$. The values of $K$ and $\Omega_r$ are also plotted. The first observation that can be done is that for all $\gamma$, $K$ is comprised between $\tik_b$ and $\tik_c$. Hence, the maximum of $r$ always occurs for a wave that is stable without piezoelectric damping and is destabilized by addition of piezoelectric damping.

Let us focus at first on the behavior of $R$ for $\gamma \rightarrow \infty$. One is tempted to conclude that one has to choose a large value of $\gamma$ to optimize the energy conversion. But it appears on Figs.~\ref{fig:R_gamma}(b)--(c) that $K \rightarrow \tik_c$ and $\Omega_r \rightarrow \tio_2(\tik_c) = 0$. This means that although the energy dissipated in the electrical circuits grows with $\gamma$, the period of the oscillations diverges \change{and the growth rate tends to zero}. This situation is thus far from optimal. One can also note that $\gamma \rightarrow \infty$ corresponds to a resistance that tends to infinity, or equivalently, an open circuit. For this reason, we conclude that the optimal efficiency corresponds to the local maximum of \fig \ref{fig:R_gamma}a around $\gamma=22$ instead of $\gamma \rightarrow \infty$. Moreover, \fig~\ref{fig:R_gamma}(c) shows that in the vicinity of the maximum efficiency, the characteristic time of the electrical circuits and the characteristic time of the wave are equal, indicating that optimal efficiency also results from a synchronization between the fluid-solid and electrical systems.

\begin{figure}[h]
\centering
\begin{psfrags}
\psfrag{PSK}{$\!\!\!\!\!K_{\gamma}$}
\psfrag{PSV}{$V^*$}
\epsfig{file=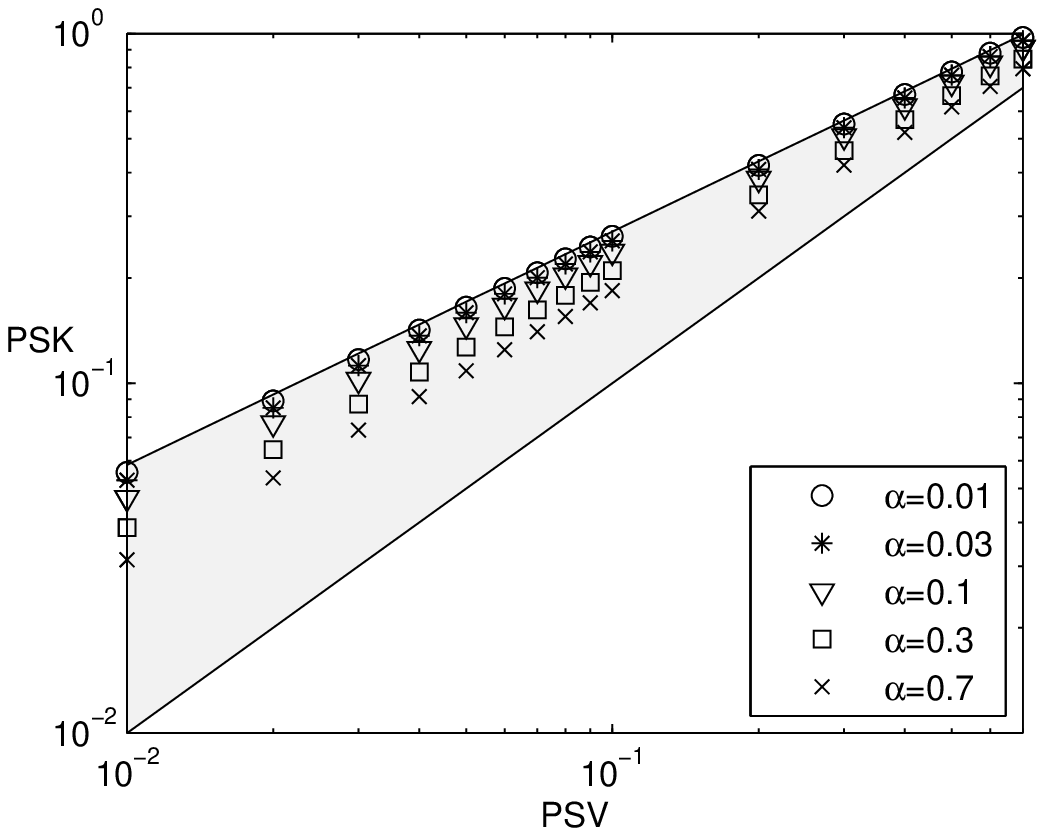,width=0.6\textwidth}
\end{psfrags}
\caption{Value of the wavenumber that maximize the efficiency, $K_{\gamma}$. Grayed region corresponds to the region of negative energy waves, $k\in[k_b,k_c]$, showing that these waves optimize the efficiency. \label{fig:koo_vstar}}
\end{figure}

For the particular values of the parameters $\alpha$ and $V^*$ used in Fig. \ref{fig:R_gamma}, negative energy waves are hence waves that optimize conversion efficiency. Let us now explore the whole $(\alpha,V^*)$ space. To do so, we introduce $\Gamma(V^*,\alpha)$, the value of $\gamma$ that maximizes $R(\alpha,V^*,\gamma)$ and $R_{\gamma}$, $K_{\gamma}$ and $\Omega_{\gamma}$ the respective values of the efficiency, the wavenumber and the frequency at this maximum. On Fig. \ref{fig:koo_vstar}, $K_{\gamma}$ is plotted as function $V^*$ for several values of $\alpha$ between 0.01 and 0.7. The range of wavenumbers destabilized by damping, $[\tik_b,\tik_c]$ appears grayed out on this plot, showing that for any values of $\alpha$ and $V^*$, optimal efficiency occurs for a wave destabilized by damping.

%When it necessary to design an optimal circuit for conversion efficiency in a particular case, the choice of the electrical quantities is done through the optimization of $\gamma$. In order to obtain an approximated behavior of the optimal value of $\gamma$ when $\alpha$ and $V^*$ vary, the quantity $\Gamma^{-1}V^{*1/2}$ is plotted as function of $\alpha$ on Fig. \ref{fig:goo_alpha}. For values of $\alpha$ up to 0.5, it appears that $\Gamma$ is well approximated by $2.5V^{*-1/2}\alpha^{-1/2}$.

\begin{figure}[h]
\centering
\begin{psfrags}
\psfrag{PSRg}{$R_{\gamma}$}
\psfrag{PSalpha}{$\alpha$}
\epsfig{file=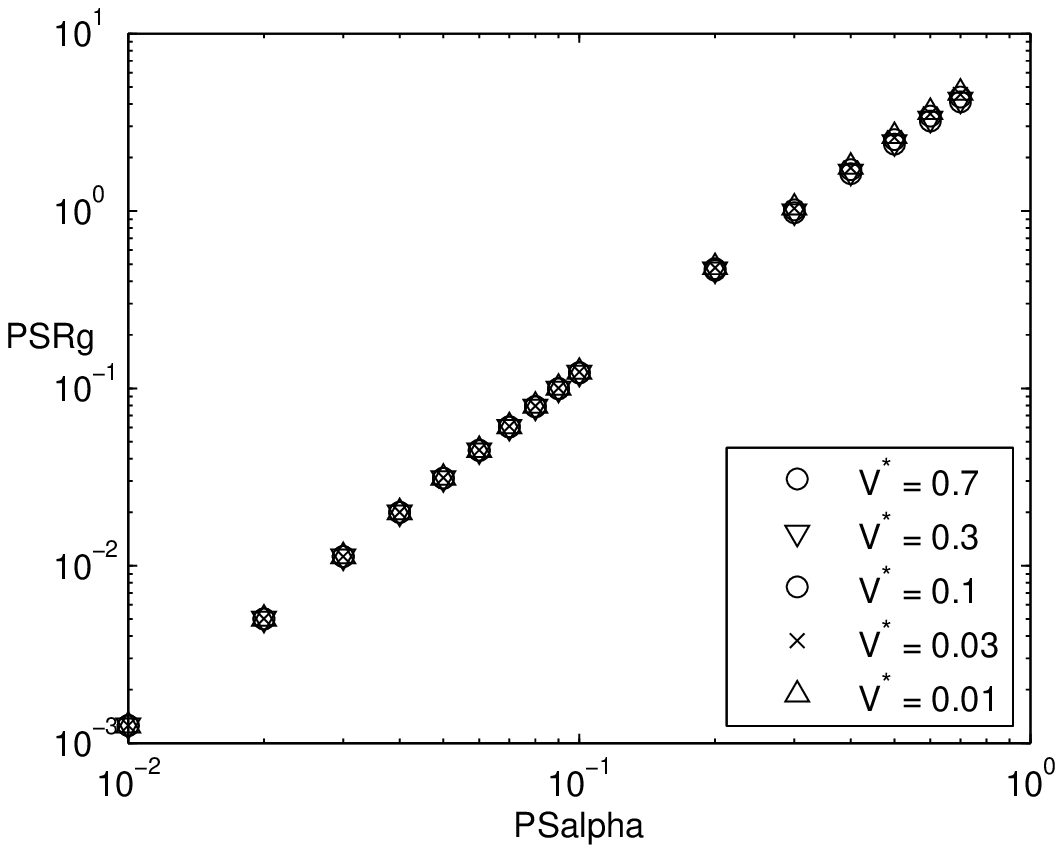,width=0.6\textwidth}
\end{psfrags}
\caption{Optimal energy conversion efficiency, $R_{\gamma}$, as function of $\alpha$ for different values of the non-dimensional velocity $V^*$. \label{fig:Rg_alpha}}
\end{figure}

Finally, the optimal efficiency $R_\gamma$ is plotted as function of $\alpha$ for different values of $V^*$ on Fig. \ref{fig:Rg_alpha}. All curves gather on a single line, indicating that the optimal efficiency scales as $\alpha^2$ and does not depend on the flow velocity. This enlightens the importance of maximizing the coupling coefficient for such application.

Before addressing the finite length system, let us summarize the main point of the local analysis conducted in this section: The particular range $[\tik_b, \tik_c]$ of wavenumbers has been emphasized, in which piezoelectric damping has a destabilizing effect and analysis of the efficiency has shown that energy conversion efficiency is maximum for waves in this range.

%%%%%%%%%%%%%%%%%%%%%%%%%%%%%%%%%%%%%%%%%%%%%%%%%%%%%%%%%%%%%%%%%%%%%%%%%
\section{Linear dynamics of a finite-length piezoelectric flexible plate}
\label{sec:global}
%%%%%%%%%%%%%%%%%%%%%%%%%%%%%%%%%%%%%%%%%%%%%%%%%%%%%%%%%%%%%%%%%%%%%%%%%

\noindent In the previous section, we focused on the local stability analysis of an infinitely long flexible plate coupled through piezo-electric patches to a purely dissipative electrical system. In this section, we are interested in the global stability analysis and conversion efficiency of a plate with finite length, coupled to the same piezo-electric dissipative system. The wave analysis approach of the local study is here replaced by a study of global modes that takes into account the boundary conditions on the finite plate and effects such as vortex shedding downstream from the solid.

\subsection{Non-dimensional equations and fluid model}
%%%%%%%%%%%%%%%%%%%%%%%%%%%%%%%%%%%%%%%%%%%%%%%%%%%%%%

\noindent The natural length scale of the problem is now $L$, the streamwise length of the plate. The non-dimensionalization of the system's equations must  be modified, and $L$ and $L/U_\infty$ are now used as characteristic length and time scales respectively. Pressure, charge densities and voltage remain non-dimensionalized by $\rho U_\infty^2$, $U_\infty\sqrt{\mu c}$ and $U_\infty\sqrt{\mu/c}$. The non-dimensional form of a variable $a$ is noted $\hat{a}$ to avoid confusion with the infinite system's notations, so that Eqs.~\eqref{eq:coupledw2}--\eqref{eq:coupledq2} now become:
\begin{align}
\frac{1}{U^{*2}}(1+\alpha^2)\hat{w}''''+\ddot{\hat{w}}-\frac{\alpha}{U^*}\hat{q}''&=-M^*\hat{p},\label{eq:dim1}\\
\beta\dot{\hat{q}}+\hat{q}-\frac{\alpha}{U^*}\hat{w}''&=0,\label{eq:dim2}
\end{align}
where the coupling parameter $\alpha$ was defined in Eq.~\eqref{eq:alpha} and
\begin{equation}\label{coeffs}
M^*=\frac{\rho_f L}{\mu},\quad\beta=\frac{cU_\infty}{gL}=\frac{\gamma}{M^*},\quad U^*=UL\sqrt{\frac{\mu}{B}}=V^*M^*.
\end{equation}
The voltage induced on the piezoelectric system is in non-dimensional form:
\begin{equation}\label{eq:tension}
\hat{v}=\hat{q}-\frac{\alpha}{U^*}\hat{w}''.
\end{equation}
The parameters $M^*$ and $U^*$ are the classical mass ratio and non-dimensional velocity used in previous flag stability analysis \citep{eloy2007,michelin2009d}. $\beta$ is the non-dimensional characteristic time-scale of the output circuit, relative to the advective time-scale $L/U_\infty$ chosen as a reference in this section. 

$M^*$ can also be interpreted as a length ratio $L/\eta$ where $\eta=\mu/\rho_f$ is the characteristic length scale introduced in Section \ref{sec:local}. The parameters $U^*$ and $\beta$ are the finite-system equivalent to the $V^*$ and $\gamma$ parameters of the local analysis [see Eqs.~\eqref{eq:Vstar} and \eqref{eq:gamma}]. In the limit of a long flexible plate, $M^*\gg 1$, the local dynamics studied in the previous section is expected to become dominant over finite-length effects such as the influence of the wake or boundary conditions.

Clamped-free boundary conditions are imposed on the finite-length flexible plate: 
\begin{align}
\textrm{    for   } \hat{x}=0\qquad&\left\{\begin{array}{l}\hat{w}=0\\\hat{w}'=0\end{array}\right.\label{eq:bcdim1}\\
\textrm{    for   } \hat{x}=1\qquad&\left\{\begin{array}{l}(1+\alpha^2)\hat{w}''-\alpha U^*\hat{q}=0\\(1+\alpha^2)\hat{w}'''-\alpha U^*\hat{q}'=0\end{array}\right..\label{eq:bcdim2}
\end{align}
Note that, Eq.~\eqref{eq:bcdim2} expresses the vanishing at the free end of the total internal torque and normal stress.

The solution of the linearized equations Eqs.~\eqref{eq:dim1}--\eqref{eq:dim2} is sought in the form of global modes
\begin{equation}\label{eq:norm_modes}
\left(\begin{array}{c} \hat{w}\\ \hat{q}\end{array}\right)=\Real\left[\left(\begin{array}{c} W(x)\\Q(x)\end{array}\right)\ee^{-\ci\hat\omega t}\right]
\end{equation}
 where $\hat\omega$ can be complex. Similarly, $V(x)$ can be defined from the potential $\hat{v}$ and is easily obtained from Eq.~\eqref{eq:tension}.
 
%%%%%%%%%%%%%%%%%%%%%%%%%%%%%%%%%%%%%%%%%%%%%%%%%%%%%%%%
\subsection{Pressure forcing on the finite-length plate}
%%%%%%%%%%%%%%%%%%%%%%%%%%%%%%%%%%%%%%%%%%%%%%%%%%%%%%%%
\noindent The pressure forcing on the flexible plate is computed from a potential flow approximation. In the case of a finite-length solid, we must account for the shedding of a vortex wake from the trailing edge of the plate. We follow here the double-wake method \citep{guo2000,eloy2007,michelin2009d}: The flow around the plate is potential everywhere except on the horizontal axis. On the plate, a bound vorticity distribution is present to satisfy the no-normal flow boundary condition on the plate's upper and lower surfaces. The free wake vorticity is not solved for explicitely as in the Vortex Sheet approach \citep{kornecki1976, michelin2009c} but it is instead imposed that the pressure discontinuity across the plate must vanish at both ends of the flexible plate ($[\hat{p}](0,t)=[\hat{p}](1,t)=0$). At the trailing edge, this is consistent with the shedding of a free horizontal vortex sheet that can not sustain any pressure force. At the leading edge, it introduces an ``upstream wake" whose physical origin is debatable. However, this method has been shown to predict the instability threshold and growth-rates correctly, particularly for intermediate-to-large mass ratio $M^*$ where the instability threshold and mode structures predicted by both methods are very similar \citep{michelin2009d}.

Using Fourier decomposition in the axial direction, it can be shown that the longitudinal gradient of the pressure jump $\partial [\hat{p}]/\partial x$ satisfies the following singular integral equation \citep{eloy2007}:
\begin{equation}
\frac{1}{2\pi}\int_0^1\pard{[\hat{p}]}{\xi}(\xi,t)\frac{\dd\xi}{x-\xi}=\left(\pard{}{t}+\pard{}{x}\right)^2\hat{w}(x,t),\qquad [\hat{p}](1,t)=[\hat{p}](0,t)=0.
\end{equation}
This singular equation can be formally solved for $\partial [\hat{p}]/\partial x$, and after integration of the pressure gradient from the leading edge, the pressure forcing on the plate is obtained as a function of the plate's displacement $\hat{w}(x,t)$. Using the normal mode decomposition in Eq.~\eqref{eq:norm_modes}, for a given mode shape $W(x)$, the pressure jump $P(x)$ can be decomposed as
\begin{equation}
P(x)=-\hat\omega^2 P^M(x)-2\ci\hat\omega P^G(x)+P^K(x)
\end{equation}
with
\begin{equation}
\frac{1}{2\pi}\int_0^1\pard{}{\xi}\left(\begin{array}{c}P^{K}[W]\\P^{G}[W]\\P^{M}[W]\end{array}\right)\frac{\dd\xi}{x-\xi}=\left(\begin{array}{c}W''(x)\\W'(x)\\W(x)\end{array}\right),\qquad
 \left(\begin{array}{c}P^{K}[W]\\P^{G}[W]\\P^{M}[W]\end{array}\right)(x=0,1)=0.\label{eq:pressureeq}
\end{equation}
$P^M$, $P^G$ and $P^K$ represent added inertia, gyroscopic and added stiffness effects on the particular mode considered, and can be expressed as linear operators on the mode shape $W$ by inverting Eq.~\eqref{eq:pressureeq}. The system \eqref{eq:dim1}--\eqref{eq:dim2} can then be rewritten as

\begin{align}
-\hat\omega^2\left(W+P^M[W]\right)-2\ci\hat\omega P^G[W]+\frac{(1+\alpha^2)}{U^{*2}}\left(W''''\right.&\left.+P^K[W]\right)-\frac{\alpha}{U^*}Q''=0\label{eq:adim1}\\
(-\ci\hat\omega\beta+1)Q-\frac{\alpha}{U^*}W''&=0,\label{eq:adim2}
\end{align}
and boundary conditions are readily obtained from Eqs.~\eqref{eq:bcdim1}--\eqref{eq:bcdim2}.

For given values of the four non-dimensional parameters $\alpha$, $\beta$, $U^*$ and $M^*$, discretizing $W(x)$ and $Q(x)$ on the first $N$ Chebyshev Gauss-Lobatto points, the above system together with boundary conditions in Eq.~\eqref{eq:bcdim1}--\eqref{eq:bcdim2}, can be written using a collocation method as an eigenvalue problem for the vector $[W,-\ci\hat\omega W,Q]^T$. 

%%%%%%%%%%%%%%%%%%%%%%%%%%%%%%%%%%%%%%%%%%%%%%%%%%%%%%%%%%%%%%%%%%%%%
\subsection{Stability analysis and impact of piezo-electric coupling}
%%%%%%%%%%%%%%%%%%%%%%%%%%%%%%%%%%%%%%%%%%%%%%%%%%%%%%%%%%%%%%%%%%%%%
\noindent The flexible plate in axial flow classically becomes unstable to fluttering above a critical velocity ratio $U^*_\textrm{crit}$ that depends on the fluid-solid mass ratio $M^*$ [e.g. \cite{eloy2007}]. We consider here the evolution of this stability threshold with the piezo-electric coupling $\alpha$ and the circuit's response time-scale $\beta$.

\begin{figure}
\begin{center}
\includegraphics[width=0.6\textwidth]{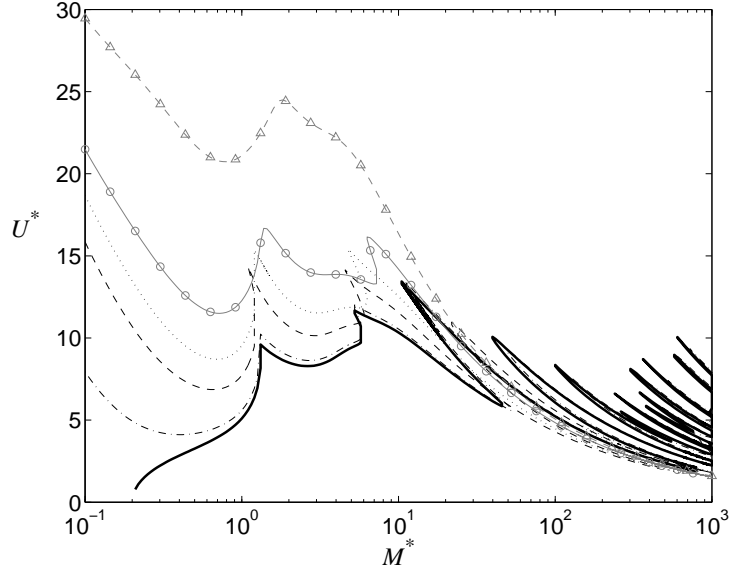}
\caption{Evolution of the stability threshold with $M^*$ for $\beta=1$ and $\alpha=0$ (thick black), $\alpha=0.2$ (dash-dotted), $\alpha=0.5$ (dashed), $\alpha=0.7$ (dotted), $\alpha=1$ (solid-circle) and $\alpha=2$ (solid-triangles). $\alpha=0$ corresponds to the uncoupled flexible plate without the piezoelectric system.}\label{fig:seuil_b1}
\end{center}
\end{figure}

For a fixed $\beta$, increasing $\alpha$ enhances the coupling between the piezo-electric and mechanical systems: an additional rigidity is introduced and energy is dissipated in the resistive circuit. The overall effect of the piezo-electric coupling $\alpha$ is therefore expected to be a stabilization of the fluttering modes. Figure \ref{fig:seuil_b1} shows that this is generally the case except at large values of $M^*$ where destabilization of some modes through the piezo-electric coupling is possible. For large enough $\alpha$ however, all modes are stabilized, in agreement with physical intuition on the stabilizing effect of a dissipative system.

Nonetheless, it is important to note that the effect of $\alpha$ is highly dependent on the tuning of the system's frequency to that of the output circuit and Figure \ref{fig:seuil_b1} corresponds to a configuration where both the fluid-solid system and the electrical circuit have similar fundamental frequencies. 

\begin{figure}
\begin{center}
\subfigure[$\beta\leq 1$]{\includegraphics[width=0.6\textwidth]{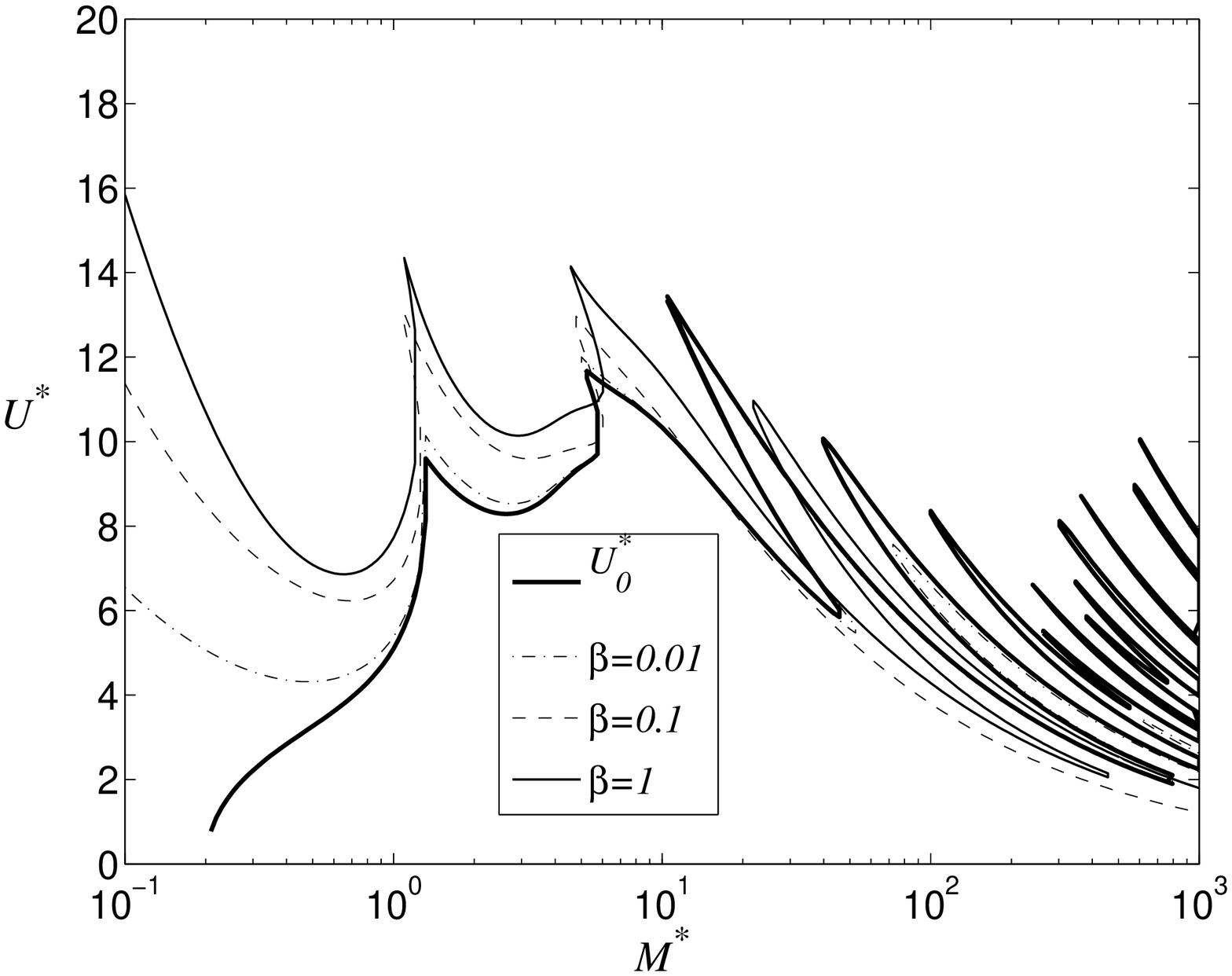}} \\
\subfigure[$\beta\geq 1$]{\includegraphics[width=0.6\textwidth]{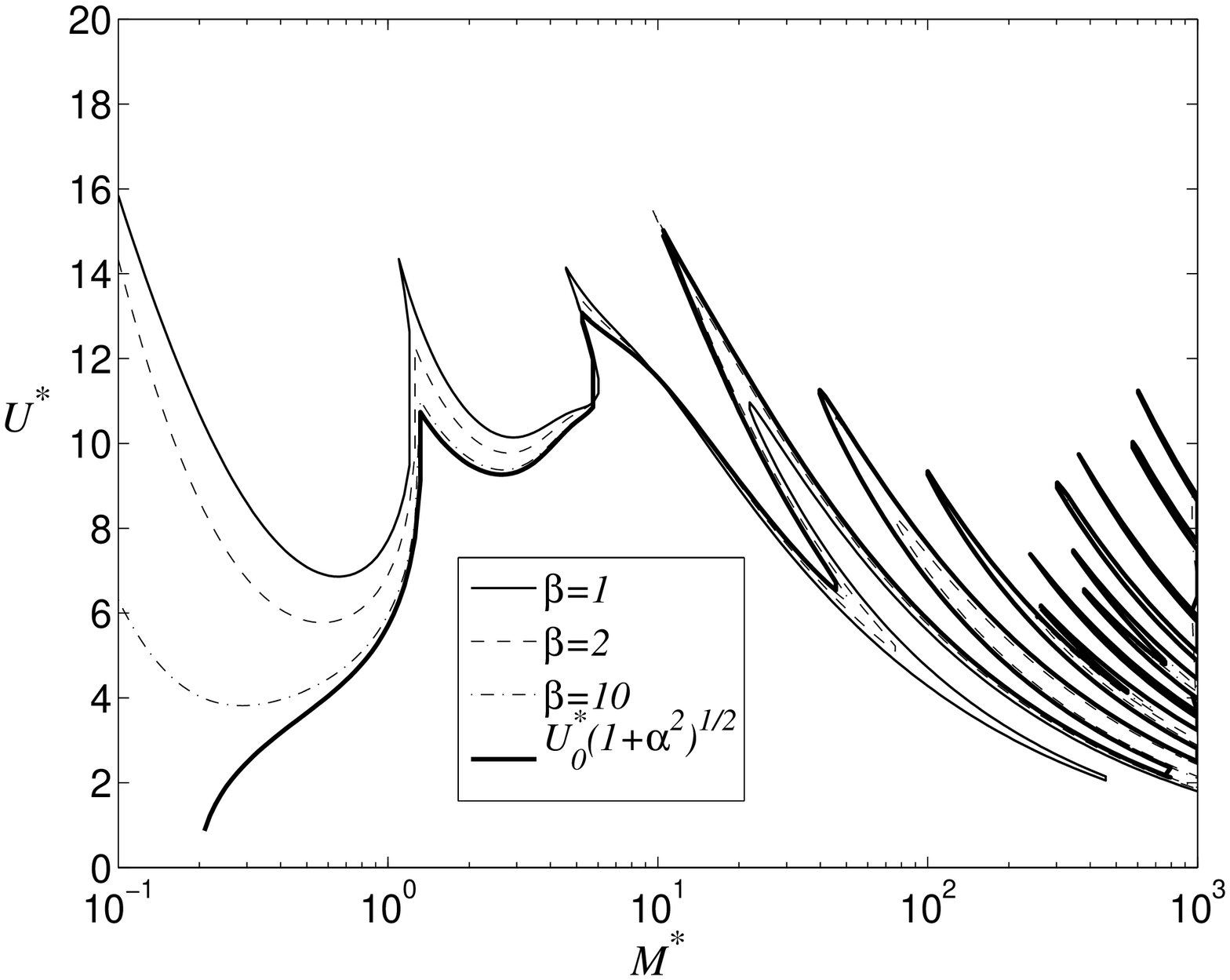}} 
\caption{Evolution of the stability threshold when $\beta$ is varied for $\alpha=0.5$. For clarity, only values of $\beta$ lower (resp. greater) than 1 are represented in (a) [resp. (b)]. In (a), the solid black line corresponds to the threshold without any coupling (no piezo-electric). In (b), the solid grey line corresponds to an uncoupled system with modified rigidity $B'=B(1+\alpha^2)$.}\label{fig:seuil_a05}
\end{center}
\end{figure}

The ratio $\beta$ is also a measure of the resistance of the electric circuit, and the limit $\beta\ll 1$ corresponds to shunted piezo-electric elements. In that case, charge transfers are instantaneous and the electrostatic balance of the piezo-electric material is achieved at all times. The electric potential difference $\hat{v}$ in the piezo-electric remains negligible, and so does the piezo-electric torque on the plate. Therefore, in the limit $\beta\ll 1$, the system behaves like the uncoupled system as observed on Figure \ref{fig:seuil_a05}(a).

In the large-$\beta$ limit, the output circuit behaves like an open loop: electric charge transfers in the circuit between the two faces of the piezo-electric material are negligible. Because of the piezo-electric coupling, the plate's bending induces an electrostatic potential difference in the piezoelectric patches that, in return, creates an additional rigidity on the plate [see Eqs.~\eqref{eq:coupledw2}--\eqref{eq:coupledq2}]. The system is therefore equivalent to a flexible plate of modified dimensional bending rigidity $B'=B(1+\alpha^2)$. This behavior is confirmed on Figure \ref{fig:seuil_a05}(b) where the stability threshold is observed to converge at large $\beta$ toward $U^*_0\sqrt{1+\alpha^2}$, with $U^*_0(M^*)$ the uncoupled stability threshold.

Between these two limits $\beta\ll 1$ and $\beta\gg 1$, the minimum velocity required for flutter to develop is observed to be significantly increased for low values of $M^*$ (heavy flags) while at large $M^*$ (typically $M^*\geq 50$), a destabilization is observed and is maximum for $\beta=O(1)$ when the dynamics of the fluid-solid system and the electrical circuit have time-scales of the same order. We also observe that for $\beta=O(1)$ the different branches observed on the stability threshold for large $M^*$ and associated with the instability region of successive high-wave number modes, disappear. Beyond the conclusions on the evolution of the stability threshold, these results emphasize the existence of a maximum in the coupling effects for values of $\beta=O(1)$.

\subsection{Conversion efficiency}
%%%%%%%%%%%%%%%%%%%%%%%%%%%%%%%%%%%

\subsubsection{Definition in the finite-length system}
%%%%%%%%%%%%%%%%%%%%%%%%%%%%%%%%%%%%%%%%%%%%%%%%%%%%%%
\noindent Total harvested energy estimates are not possible within a purely linear framework. However, the conversion efficiency (or normalized harvested energy) defined in Eq.~\eqref{eq:efficiency} provides some important information about the energy transfers in the linearly unstable modes of the coupled systems and its ability to produce electrical power from the flapping of the flexible plate. The spatial average in Eq.~\eqref{eq:efficiency} is now taken over the entire length of the plate and the conversion efficiency $r$ becomes in the non-dimensional notations of this section
\begin{equation}\label{eq:efficiency2}
r=\frac{\displaystyle\frac{1}{\beta}\int_{t}^{t+\frac{2\pi}{\hat\omega_r}}\int_0^1\hat{v}^2\dd t'\dd \hat{x}}{\displaystyle\frac{\hat\omega_r}{2\pi}\int_{t}^{t+\frac{2\pi}{\hat\omega_r}}\int_0^1\left(\frac{1}{2}\dot{\hat{w}}^2+\frac{1}{2U^{*2}}\hat{w}^{''2}+\frac{1}{2}\hat{v}^2\right)\dd t'\dd \hat{x}}\cdot
\end{equation}
For a function $f(x,t)=\Real\left(F(x)\ee^{-\ci\hat\omega t}\right)$ with $\hat\omega=\hat\omega_r+\ci\hat\omega_i$ ($f$ being one of $\hat{v}$, $\hat{q}$, $\hat{w}$...),
\begin{equation*}
\int_{t} ^{t+2\pi/\omega_r}f(x,t')^2\dd t'=\frac{(\ee^{4\pi\frac{\hat\omega_i}{\hat{\omega}_r}}-1)\,\ee^{2\hat\omega_i t}}{4\hat\omega_i}\left(|F(x)|^2+\Real\left[\frac{\ci\hat\omega_i\ee^{-\ci\hat\omega_r t}}{\hat\omega}F(x)^2\right]\right),
\end{equation*}
and Eq.~\eqref{eq:efficiency2} becomes
\begin{equation}
r(t)=\frac{\displaystyle G_1+\Real\left(\frac{\ci G_2\hat\omega_i}{\hat\omega}\ee^{2\ci \hat\omega_r t}\right)}{\displaystyle H_1+\Real\left(\frac{\ci H_2\hat\omega_i}{\hat\omega}\ee^{2\ci \hat\omega_r t}\right)},
\end{equation}
with $G_1=\int_0^1|G(x)|^2\dd x$ and $G_2=\int_0^1G(x)^2\dd x$, with
\begin{equation}
G(x)=\frac{V(x)^2}{\beta},
\end{equation}
and $H_1$ and $H_2$ are defined similarly from 
\begin{equation}
H(x)=\frac{1}{2}\left([\ci\hat\omega W(x)]^2+\frac{1}{U^{*2}}W''(x)^2+\frac{1}{2}V(x)^2\right).
\end{equation}

Unlike in the local analysis of Sec.~\ref{sec:local}, and because the different functions are not periodic on the averaging length anymore, $r$ is still a function of time, except at the stability threshold where $\hat\omega_i=0$. $r(t)$ is however periodic, and for unstable flutter modes, the fluctuations of $r$ around its mean value are weak as the mode's growth rate is generally small compared to its frequency. In the following, most of the analysis will be performed at the stability threshold, where $r$ is a constant. In all other cases, its time-average $\bar{r}$ over a period will be taken as a measure of the conversion efficiency and is easily computed from $\hat\omega$ and the mode shape functions $V$ and $W$.

\subsubsection{Energy conversion and resonance}
%%%%%%%%%%%%%%%%%%%%%%%%%%%%%%%%%%%%%%%%%%%%%%%

\begin{figure}
\begin{center}
%\begin{psfrags}
%\psfrag{PSU}{$U^*$}
%\psfrag{PSM}{$M^*$}
\includegraphics[width=0.6\textwidth]{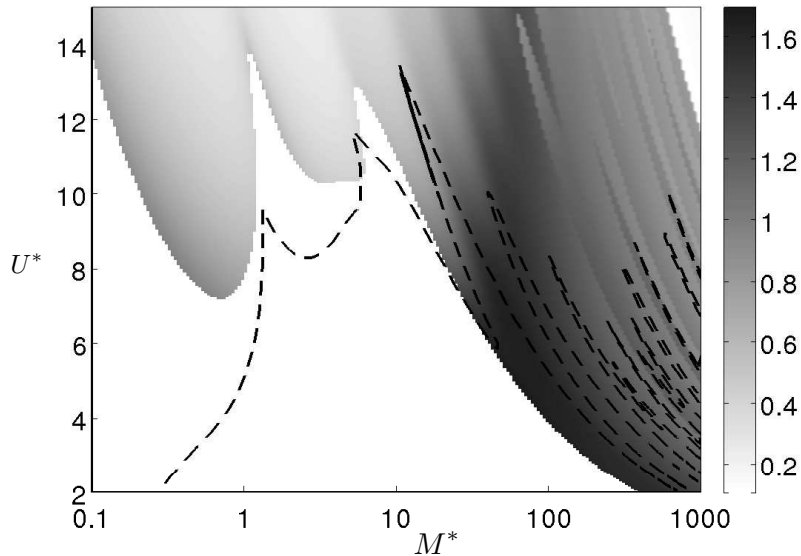}
%\end{psfrags}
\caption{Conversion efficiency of the dominant unstable mode for $\alpha=0.5$ and $\beta=0.25$. In the white regions, all the modes are stable. The stability threshold obtained for the uncoupled system ($\alpha=0$) is shown as a dashed line.}\label{fig:conv_eff_beta025a05}
\end{center}
\end{figure}

\noindent We are interested here in the coupled behavior of the fluid-solid and electric systems for a given piezoelectric material (fixed $\alpha$). Figure \ref{fig:conv_eff_beta025a05} shows the variations with $M^*$ and $U^*$ of the conversion efficiency associated with the dominant mode for $\alpha=0.5$ and $\beta=0.25$, above the stability threshold. We observe that this conversion efficiency is increased significantly for large $M^*$, corresponding to high fluid-to-solid mass ratios (in water for example) or long systems. The conversion efficiency is also observed to be maximum in the regions destabilized by the piezo-electric coupling. This is consistent with the conclusions of the local analysis, where a maximum of the conversion efficiency was observed for the Negative Energy Wave destabilized by the introduction of piezoelectric damping.

\begin{figure}
\begin{center}
%\begin{psfrags}
%\psfrag{PSB}{$\beta$}
%\psfrag{PSM}{$M^*$}
\includegraphics[width=0.6\textwidth]{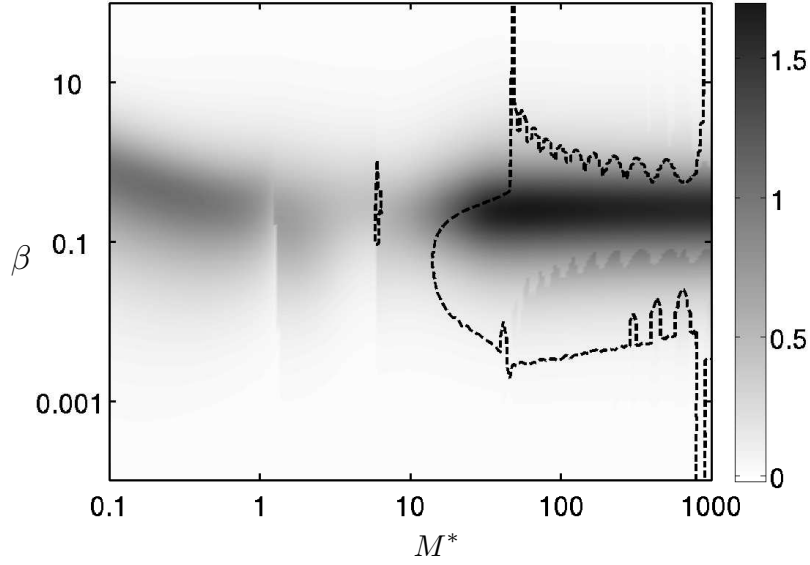}
%\end{psfrags}
\caption{Conversion efficiency $\bar{r}$ of the dominant mode (neutral) at the stability threshold for $\alpha=0.5$. The thick black line shows the limits of the regions of the $(M^*,\beta)$-plane destabilized in comparison with the uncoupled problem $\alpha=0$.}\label{fig:conv_eff_seuildisp_a05}
\end{center}
\end{figure}

Such destabilized regions are located in the vicinity of the instability threshold and we will from now on focus on this parameter region, as it is also most relevant given the linear character of this study. The map of the conversion efficiency for the dominant (and neutrally stable) mode at the threshold is shown on Figure \ref{fig:conv_eff_seuildisp_a05}. It confirms the existence of a maximum of conversion efficiency at high $M^*$ for $\beta\sim 0.1$--$1$. We also observe that the maximum efficiency and the corresponding value of $\beta$ are independent of $M^*$ for $M^*\geq 20$--$50$. This suggests that the dynamics leading to the maximum energy conversion are dominated by local effects as will be discussed further in Section \ref{sec:discussion}. Figure \ref{fig:conv_eff_seuildisp_a05} also confirms that the maximum conversion efficiency is reached when the flexible plate is destabilized by the piezo-electric coupling.

\begin{figure}
\begin{center}
\begin{tabular}{ccc}
\hspace{-.5cm}\subfigure[$M^*=1$]{\includegraphics[width=0.3\textwidth]{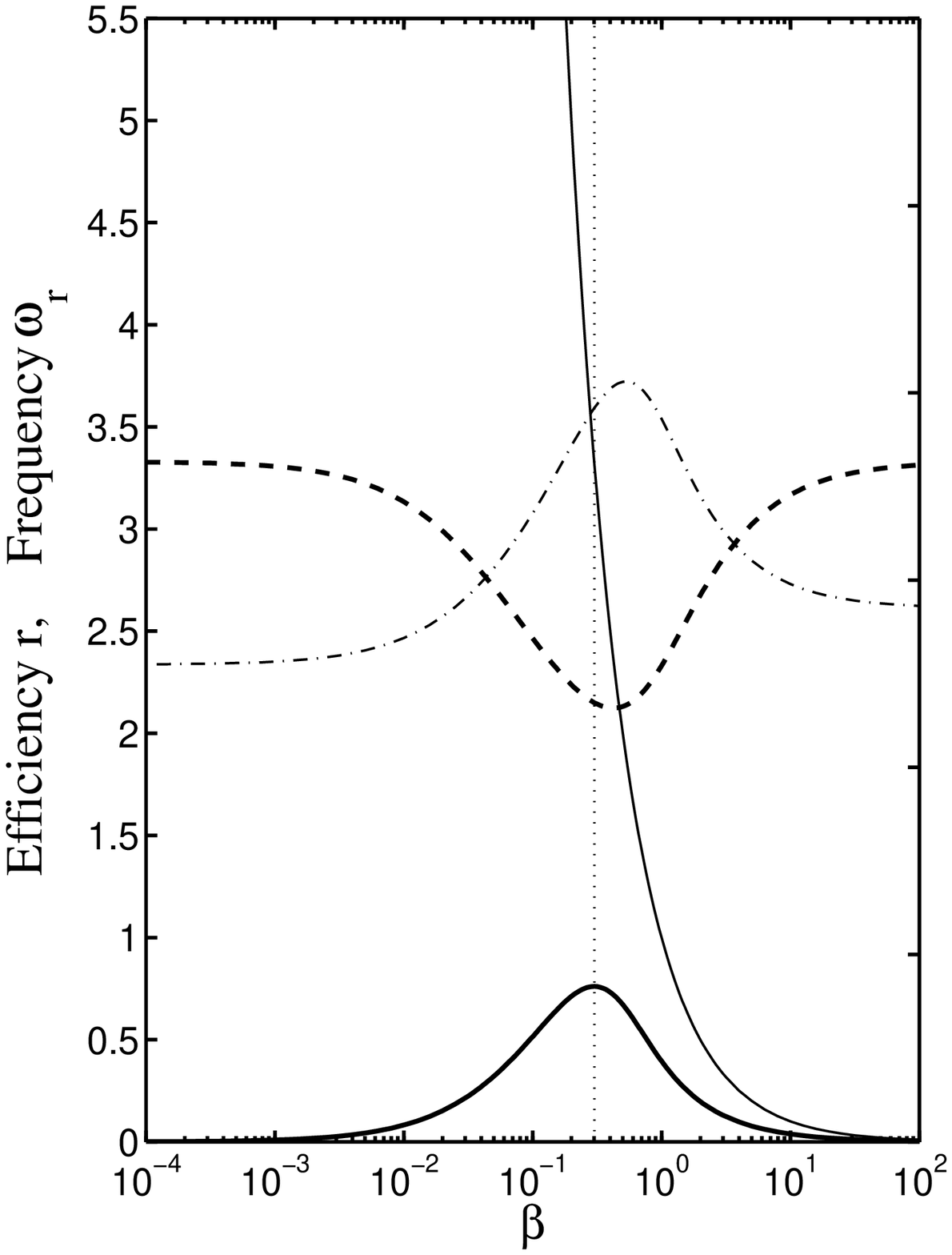}}&
\subfigure[$M^*=10$]{\includegraphics[width=0.3\textwidth]{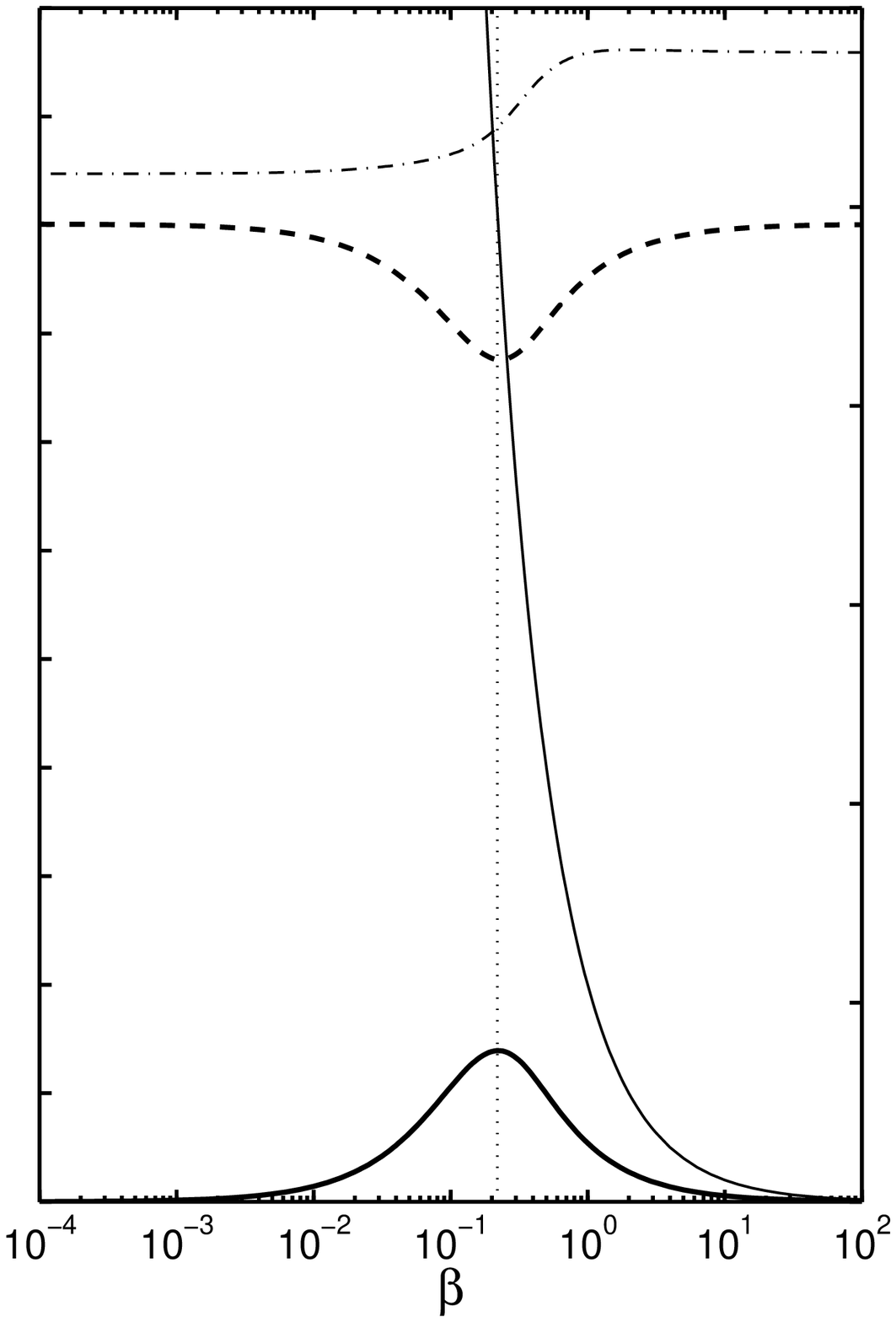}}&
\subfigure[$M^*=100$]{\includegraphics[width=0.3\textwidth]{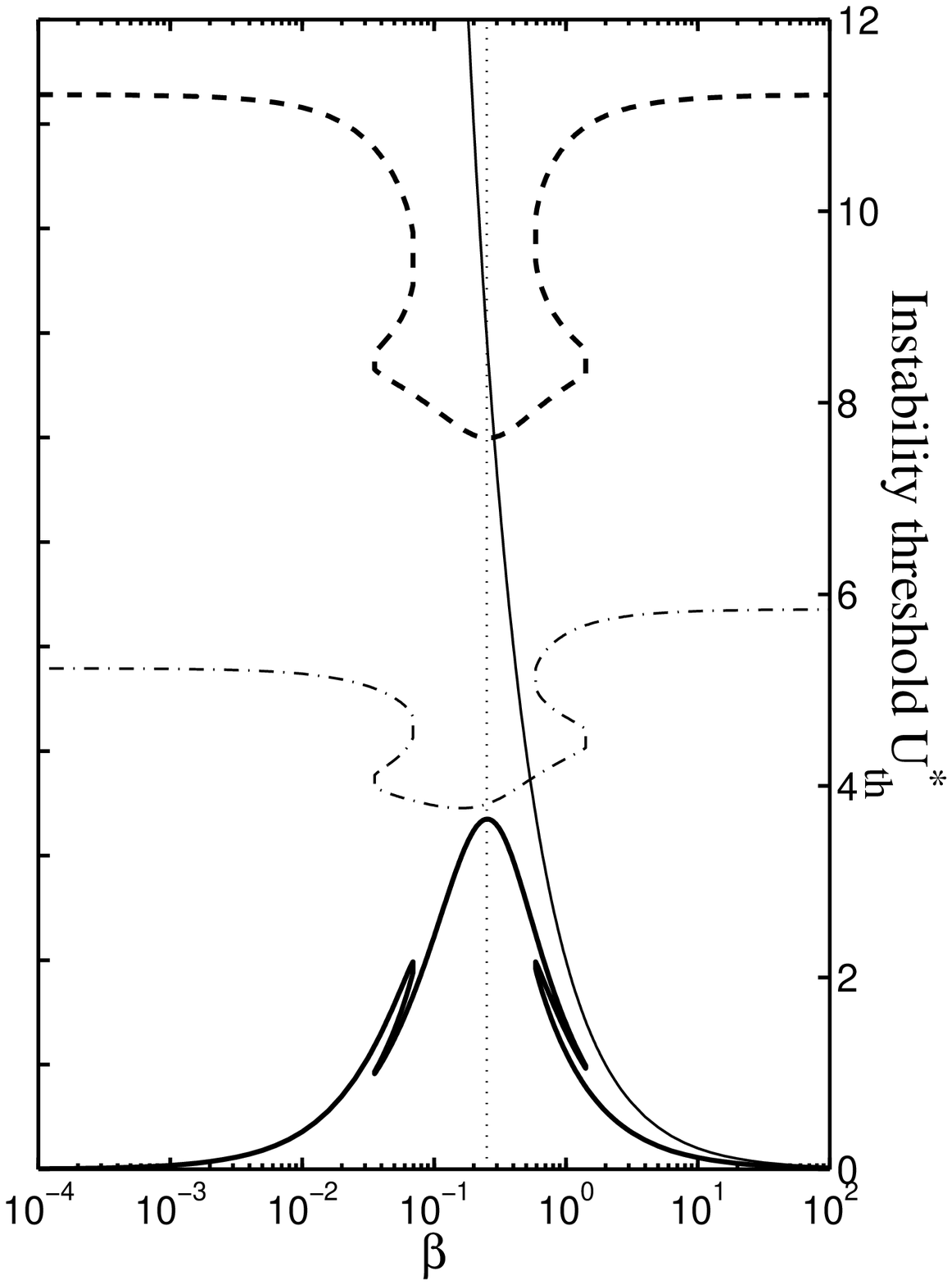}}
\end{tabular}
\caption{Evolution with $\beta$ at the stability threshold of the conversion efficiency (thick solid) and frequency (thick dashed) for $\alpha=0.5$ and $M^*=1$ (a), $M^*=10$ (b) and $M^*=100$ (c). The evolution of the critical velocity $U^*$ at the threshold is also shown on each plot (thin dash-dotted) and the thin solid line represents $1/\beta$ for reference. The vertical dotted line indicates the value of $\beta$ for which the conversion efficiency is maximum.}\label{fig:a05M100}
\end{center}
\end{figure}

In that region of the parameter space, the frequency of the dominant mode varies between $3$ and $5$ (Figure \ref{fig:a05M100}), which is of the same order of magnitude as the value of $1/\beta$ near the maximum efficiency, therefore suggesting a resonance-type phenomenon between the frequency of the neutrally stable mode and the characteristic frequency of the output electric system. The existence of a maximum for the conversion efficiency for $\beta\omega_r\simeq 1$ is confirmed on Figure \ref{fig:a05M100}. Note that the agreement is excellent for $M^*=100$ while it is not as good for $M^*=1$ which lies outside of the maximum efficiency region.

The evolution of the maximum efficiency with $\beta$ on Figure \ref{fig:a05M100} is highly reminiscent of the variations of the dissipated energy in the resistive part of a $RC$-electrical circuit forced by a generator at a given frequency $\omega$. The role of the forcing generator is played here by the fluid-solid system that imposes charge transfers and non-zero potential in the piezo-electric material through the deformation of the flexible plate [Eq.~\eqref{eq:coupledq2}]. The forcing frequency is however not independent from the output circuit, as shown on Figure \ref{fig:a05M100} where we observe a significant variation with $\beta$ of the frequency of the dominant mode. This modification of the forcing frequency is a result of the piezoelectric feedback of the electrical system on the fluttering dynamics. In that regard, the observed efficiency peak differs from a traditional resonance and corresponds to a tuning of the fluid-solid frequency to the output system's when the two time-scales are of the same order.

The comparison of the three cases $M^*=1$, $M^*=10$ and $M^*=100$ also confirms that the maximum conversion efficiency is significantly larger when destabilization by damping occurs (lower value of the threshold when $\beta\omega_r\simeq 1$).

\subsubsection{Influence of the piezoelectric coupling}
\begin{figure}
\begin{center}
\includegraphics[width=0.6\textwidth]{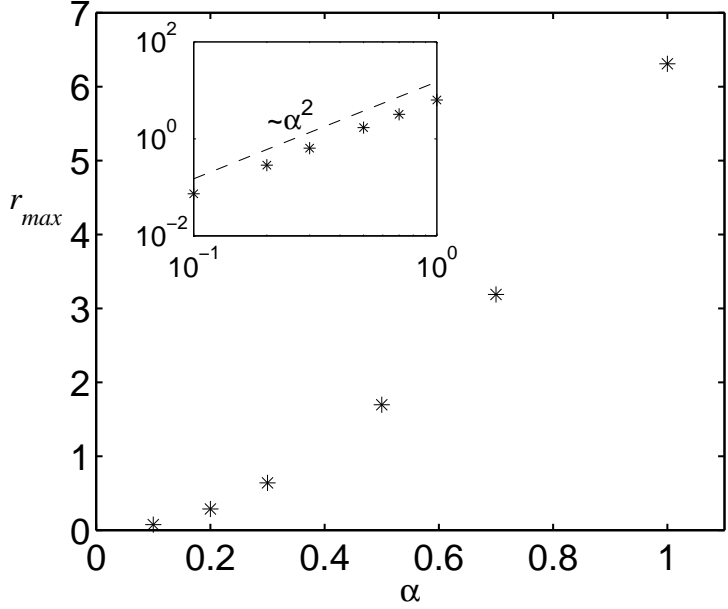}
\caption{Evolution of the maximum conversion efficiency $r_\textrm{max}$ with the coupling coefficient $\alpha$. $r_\textrm{max}$ is defined as the maximum of $r$ evaluated at the stability threshold for all values of $\beta$ and $M^*$. The insert plot represents the same quantities in log-log format, showing that $r_\textrm{max}\sim\alpha^2$.}\label{fig:alpha_dependence}
\end{center}
\end{figure}

%%%%%%%%%%%%%%%%%%%%%%%%%%%%%%%%%%%%%%%%%%%%%%%%%%%%%%%
\noindent In Figure \ref{fig:conv_eff_seuildisp_a05}, we observed that the conversion efficiency reaches a maximum value of $1.7$ for large $M^*$ and $\beta\omega_r\simeq 1$. Figure \ref{fig:alpha_dependence} shows that the maximum conversion efficiency at the stability threshold is strongly influenced by the coupling coefficient $\alpha$, with a clear scaling $r_\textrm{max}\sim\alpha^2$, and this confirms the discussion of Section \ref{sec:local}. It is not surprising to obtain values of $r_\textrm{max}$ greater than one as it is not a thermodynamic efficiency but a measure of the harvested energy over a period relative to the average stored energy over the same period in the solid and capacitative systems. Note also, that unlike $\beta$, $U^*$ and $M^*$, that are determined by the tuning of the output system properties, of the plate's length and inertia and of the flow conditions, $\alpha$ only depends on the characteristics of the piezo-electric material and the plate's rigidity. For a given piezo-electric system $\alpha$ is fixed and Figure \ref{fig:alpha_dependence} provides an upper bound of the achievable conversion efficiency.

%%%%%%%%%%%%%%%%%%%%%%%%%%%%%%%%%
\section{Discussion}
\label{sec:discussion}
%%%%%%%%%%%%%%%%%%%%%%%%%%%%%%%%%

\noindent The fluttering of a flexible plate in an axial flow is an attractive candidate for flow energy harvesting as it can produce self-sustained periodic vibrations of the solid body. Understanding the impact of the harvesting-induced damping on the fluid-solid dynamics is an essential element to assess the potential performance of such systems. Thus the present paper investigated the linear dynamics of the fluttering flexible plate fully coupled to a simple dissipative electrical circuit through piezoelectric layers, by studying the coupling effect on the local and global instabilities as well as on the energy conversion efficiency.

The ability to quantify the total harvested energy is intrinsically limited by the linear framework considered, as saturated flapping amplitudes are not computed. However, the linear analysis is an important step to obtain some insight on the modification of the system properties, in particular its stability threshold, by the introduction of a realistic coupling to an output electrical network. Hence, it was shown that destabilization by damping can occur in such systems. The local analysis showed that the destabilized modes correspond to the so-called Negative Energy Waves of the un-damped system. This was confirmed in the global analysis by the destabilization of plates with high mass ratio where local effects are expected to dominate. In particular, we observed on Figure \ref{fig:conv_eff_seuildisp_a05} that above a certain value of $M^*$ of the order of $100$, the dependence with $M^*$ of the system properties at the stability threshold becomes negligible. 

The parameter $M^*$ can be seen as a relative measure of the length of the system to the local characteristic lengthscale. For $M^*\gg 1$, one expects the stability and properties of the dominant modes to be driven by local phenomena rather than global ones.  For given $\alpha$ and $\beta$, the critical velocity threshold remains finite which is consistent with the observation that the system is unstable locally for all $V^*=U^*M^*$. The presence of a maximum in conversion efficiency for $\beta\hat\omega_r\simeq 1$ can be identified to the maximum efficiency observed in local analysis for $\gamma\tilde\omega_r\simeq 1$ as $\beta=\gamma/M^*$ and $\hat\omega_r=\tilde\omega_r M^*$.

\change{From a practical point of view, the present results suggest that the energy conversion is more efficient when the mass ratio $M^*$ is large. This is achieved when the plate's length is long, but also when the fluid inertia is large compared to the solid's, as in water flows for example. These results also emphasize the importance of the output circuit in the energy transfers: a careful tuning of the circuit characteristic time-scale to that of the fluid-solid oscillations significantly increases the conversion efficiency from the solid to the electric system.}

Using both local and global analyses, the maximum energy conversion efficiency $r_\textrm{max}$ was found to scale as $\alpha^2$. This scaling is expected in the limit of weak coupling: if one neglects the piezo-electric coupling term in Eq.~\eqref{eq:localadimfluid1} or Eq.~\eqref{eq:dim1}, the fluid-system can be considered as only weakly modified and act as a constant forcing on the electric system. The forcing in Eq.~\eqref{eq:localadimfluid2} and Eq.~\eqref{eq:dim2} scales linearly with $\alpha$ and so is expected to scale the charge density. The harvested energy, a quadratic function of the charge density, is therefore expected to scale as $\alpha^2$. It is however surprising to observe that this scaling remains valid in a greater range of coupling. \change{This result enlightens the essential role of the coupling coefficient $\alpha$ in energy harvesting applications. This coupling coefficient can be obtained from Eqs.~\eqref{eq:rigidity} and \eqref{eq:alpha} by separating the properties of the piezoelectric materials from the geometric effects of the relative thickness of the plate and piezoelectric patches: 
\begin{equation}
\alpha=e_{31}\sqrt{\frac{1-\nu_p^2}{\varepsilon E_p}} \,\,\mathscr{G}\left(\frac{h_0}{h_p},\frac{E_0(1-\nu_p^2)}{E_p(1-\nu_0^2)}\right),
\end{equation}
where $\mathscr{G}$ is a non-dimensional function of the thickness and bending stiffness ratios. To increase the piezoelectric coupling it is therefore important to maximize $e_{31}$ or minimize the permittivity $\varepsilon$. It must be noted that changing $\varepsilon$ also impacts the capacity of the piezoelectric material, which in turn modifies the choice of the optimal energy harvesting electrical circuit through the optimization of the parameter $\beta$. Finally, once the materials are chosen, an optimal thickness ratio $h_0/h_p$ can also be obtained.
}
To get an idea of the value of \change{$\alpha$ in practical applications}, two cases may be considered that are representative of typical values found in the literature: the former consists in a mylar plate ($h_0=100\mu$m, $E_0=4$GPa) with two PVDF piezoelectric layers ($h_p=40\mu$m, $E_p=2.5$GPa, $e_{31}=0.023$C/m$^2$, $\varepsilon_{r}=11.5$), leading to an approximate value of $\alpha \sim 0.03$. The latter consists in a steel plate ($h_0=300\mu$m, $E_0=200$GPa) with two PZT piezoelectric layers ($h_p=300\mu$m, $E_p=60$GPa, $e_{31}=10$C/m$^2$, $\varepsilon_{r}=2000$) and in that case, the coupling coefficient is $\alpha \sim 0.3$. Careful design of the system is expected to further increase $\alpha$, therefore the typical value of $\alpha=0.5$ chosen in the present paper is realistic but corresponds to an upper bound estimate of the system's performance.\\

Natural extensions of this work include the study of passive resonant \citep{hagood1991} or active circuits \citep{chen2010} to determine potential improvements in the energy transfer from the electrical design. In this paper, a continuous distribution of piezoelectrics has been considered and studying the impact of the finite-length of piezoelectric patches would provide important information on the actual design of energy harvesters. Finally, as emphasized throughout the present paper, further investigation of the energy harvesting potential must include the representation of non-linear effects in the fluid and solid dynamics, to obtain the amplitude of the self-sustained oscillations of the system. Both local \citep{peake2001} and global \citep{michelin2008} analyses of the nonlinear regime have been performed for compliant panels or plates placed in an axial flow. The non-linear behavior of piezoelectric materials in the energy harvesting context has been investigated by \cite{triplett2009}. The study of the saturated regime in the fluttering dynamics of the fluid-structure-electrical fully-coupled system will provide a quantitative assessment of the actual harvested power and is the focus of subsequent work to this linear analysis.

%\section*{References}

\bibliographystyle{unsrt}
\bibliography{refs-paper}

\newpage

%\newpage
%
%\begin{figure}[h]
%\centering
%\epsfig{file=PePf_gamm15alph5e-1UR5e-2.eps,width=0.8\textwidth} \\
%\caption{Ratio between the power dissipated in the electric networks and the work done by the fluid on the structure. Dashed line indicates a negative ratio.\label{fig:powerk}}
%\end{figure}

\newpage

\newpage

\newpage
\pagebreak

\newpage
\pagebreak

\newpage
\pagebreak

\newpage
\pagebreak

\newpage
\pagebreak

\newpage
\pagebreak

\end{document}